\newcommand{\rootpath}{.}

\documentclass[12pt, draftclsnofoot, onecolumn]{IEEEtran}

\ifCLASSINFOpdf
\else
\fi
\input{head}

\newboolean{isdouble}
\setboolean{isdouble}{false} 

\begin{document}

%

\title{Disentangled Representation Learning for RF~Fingerprint Extraction under Unknown Channel Statistics}
%
%
%

\ifthenelse{\boolean{isdouble}}{
\author{Renjie Xie,~\IEEEmembership{Graduated Student Member,~IEEE},
        Wei Xu,~\IEEEmembership{Senior Member,~IEEE}, Jiabao Yu, \\
        Aiqun Hu,~\IEEEmembership{Senior Member,~IEEE}, 
        Derrick Wing Kwan Ng,~\IEEEmembership{Fellow,~IEEE}, \\
        and A. Lee Swindlehurst,~\IEEEmembership{Fellow,~IEEE}
        %
\thanks{R. Xie, W. Xu, J. Y, and A. Hu are with the National Mobile Communications Research Laboratory, Southeast University, Nanjing 210096, China (e-mail: renjie\_xie@seu.edu.cn, wxu@seu.edu.cn, yujiabao@seu.edu.cn, aqhu@seu.edu.cn).}%
\thanks{D. W. K. Ng is with the School of Electrical Engineering and Telecommunications, University of New South Wales, Sydney, NSW 2052, Australia (e-mail: w.k.ng@unsw.edu.au).}
\thanks{A. L. Swindlehurst is with the Center for Pervasive Communications and Computing, University of California, Irvine, CA 92697-2625 USA (e-mail: swindle@uci.edu).}%
}%
}{
        
\author{Renjie Xie, Wei Xu,~\IEEEmembership{Senior Member,~IEEE}, Jiabao Yu, \\
Aiqun Hu,~\IEEEmembership{Senior Member,~IEEE}, Derrick Wing Kwan Ng,~\IEEEmembership{Fellow,~IEEE},\\
 and A. Lee Swindlehurst,~\IEEEmembership{Fellow,~IEEE}
        %
\thanks{R. Xie, W. Xu, J. Y, and A. Hu are with the National Mobile Communications Research Laboratory, Southeast University, Nanjing 210096, China (e-mail: renjie\_xie@seu.edu.cn, wxu@seu.edu.cn, yujiabao@seu.edu.cn, aqhu@seu.edu.cn).}%
\thanks{D. W. K. Ng is with the School of Electrical Engineering and Telecommunications, University of New South Wales, Sydney, NSW 2052, Australia (e-mail: w.k.ng@unsw.edu.au).}
\thanks{A. L. Swindlehurst is with the Center for Pervasive Communications and Computing, University of California, Irvine, CA 92697-2625 USA (e-mail: swindle@uci.edu).}%
}%
}
\maketitle
\ifthenelse{\boolean{isdouble}}{}{\vspace{-2cm}}
\begin{abstract}

Deep learning (DL) applied to a device's radio-frequency fingerprint~(RFF) has attracted significant attention in physical-layer authentication due to its extraordinary classification performance. 
Conventional DL-RFF techniques are trained by adopting maximum likelihood estimation~(MLE). \bflag{Although their discriminability has recently been extended to unknown devices in open-set scenarios, they} still tend to overfit the channel statistics embedded in the training dataset. This restricts their practical applications as it is challenging to collect sufficient training data capturing the characteristics of all possible wireless channel environments. To address this challenge, we propose a DL framework of disentangled representation~\bflag{(DR)} learning that first learns to factor the signals into a device-relevant component and a device-irrelevant component via adversarial learning. Then, it shuffles these two parts within a dataset for implicit data augmentation, which imposes a strong regularization on RFF extractor learning to avoid the possible overfitting of device-irrelevant channel statistics, without collecting additional data from unknown channels. 
Experiments validate that the proposed approach, referred to as \bflag{DR-based RFF}, outperforms conventional methods in terms of generalizability to \bflag{unknown devices even under} unknown complicated propagation environments, e.g., dispersive multipath fading channels, even though all the training data are collected in a simple environment with dominated direct line-of-sight~(LoS) propagation paths.

\end{abstract}


%
\IEEEpeerreviewmaketitle

\section{Introduction}
\IEEEPARstart{S}{tate}-of-the-art authentication using inherent physical-layer characteristics has shown great potential in securing communication in future Internet-of-Things~(IoT) networks~\cite{peng2019deep}. Compared to conventional higher-layer authentication techniques, physical-layer authentication~(PLA) exploits inherent unique hardware  characteristics of individual devices, known as their radio-frequency fingerprint~(RFF), to perform effective authentication. Analogous to human fingerprints, these hardware characteristics are naturally caused by manufacturing deviations and are difficult to modify or tamper with~\cite{wang2016wireless}. Authentication with RFF has the advantages of short latency, low power consumption, and marginal computational overhead, which is appealing for practical implementations~\cite{hou2014physical}. 

\bflag{Generally, the hardware characteristics, i.e., RFF, are present in the transmitted signals, which arrive at the receiver over a wireless channel~\cite{danev2012physical}.}
To enable RFF authentication, it is essential to extract discriminative RFFs from the signals sent by the devices of interest \bflag{while avoiding the effects of the channel.} As a result, intensive efforts have been devoted to extracting stable RFFs. Conventionally, handcrafted features based on expert knowledge have been adopted to extract the RFFs~\cite{brik2008wireless, hall2004enhancing, knox2010agc}. However, due to the limitations of expert knowledge {of these nonlinear hardware characteristics}, handcrafted features usually suffer from low discrimination ability, and cannot cope with growing IoT applications involving massive numbers of devices~\cite{chen2020massive}.

\subsection{Related works}
\paragraph{Deep learning-based RF fingerprinting}
Deep learning~(DL)-based methods have recently been exploited for effective nonlinear feature extraction in physical-layer applications~\cite{xu2022edge, yin2022deep}, \bflag{especially for RFF extraction~\cite{merchant2018deep, yu2019robust, peng2019deep, sankhe2019oracle, sankhe2019no, ding2018specific, soltani2020rf, jian2020deep, reus2020trust, zhao2018classification, restuccia2019deepradioid, liu2020zero, xie2021generalizable, shen2022towards, hanna2020open}. In particular, the pioneering work in~\cite{merchant2018deep} first adopted a convolutional neural network (CNN) to extract RFFs from the received signals. In~\cite{yu2019robust}, the authors extended the method in~\cite{merchant2018deep} for extracting RFFs from the received signals with multiple sampling rates. The authors in~\cite{peng2019deep} proposed to use a CNN to extract RFFs after transforming the signal into a differential constellation trace figure~(DCTF). The work of~\cite{jian2020deep} investigated the classification performance of CNNs on large-scale wireless devices. }
\bflag{Thanks to the ability of DNNs to learn the RFF features themselves, the classification performance of these DL-based RFFs is noticeably better than that obtained with handcrafted RFFs. }
\bflag{Despite improvement, the above DL-based RFFs still regard RFF authentication as a closed-set classification problem, which essentially learns a classifier via maximum likelihood~(ML) estimation on a given training dataset and evaluates its performance on the devices that were present during the training. Even though these methods perform well on known devices, some of them, e.g., \cite{merchant2018deep} and \cite{yu2019robust}, have been verified to suffer from noticeable performance degradation when unknown devices are present in the system~\cite{xie2021generalizable}. }
\bflag{Since collecting data from all possible devices is not possible, a reliable RFF authentication system should not only work well on known devices but it should also be effective for unknown devices. This is the so-called open-set physical-layer authentication problem~\cite{danev2012physical, hanna2020open, xie2021generalizable}.}

\bflag{Some works have proposed to discover new devices via outlier detection~\cite{zhao2018classification, liu2020zero} or by treating all the unseen devices as a separate class~\cite{hanna2020open}. However, these methods still require the networks to be retrained to achieve compatibility with newly added legitimate devices. This is unrealistic because retraining DNNs is computationally expensive. To address the issue of open-set RFF authentication, our previous work~\cite{xie2021generalizable} proposed a data-and-model driven preprocessing module and adapted the ML estimation to a typical metric learning task to strengthen the discrimination of the DL-based RFFs. Using this method, unknown devices can be recognized without retraining, even when encountering device aging. }


\bflag{On the other hand, from a learning perspective, even though \cite{xie2021generalizable} enables ML-based RFFs to verify/identify unknown devices, they still tend to overfit the propagation environment in the training data. }
In practice, the training data collected for RFF extraction inevitability contains both hardware characteristics and the impacts of the propagation environment. The ML-based RFFs trained using data from channels under a specific propagation environment, e.g., line-of-sight (LoS)-dominated channels, tend to overfit the resulting model, particularly features that are sensitive to the propagation environment. More importantly, the methods sometimes fail to generalize to other types of channels such as those with considerable multipath. Unfortunately, this challenge cannot simply be addressed by collecting more training data that covers all possible wireless channel environments. In real-world IoT networks, collecting data representative of all possible channel conditions is prohibitively expensive if not impossible.

\paragraph{Data augmentation}
\bflag{One possible approach for avoiding this overfitting problem in ML-based RFFs is to apply data augmentation (DA) techniques~\cite{shorten2019survey} for training the RFF extractor. Conventionally, DA is applied by imposing some handcrafted transforms on the existing data to create synthetic data, referred to as handcrafted DA. For RFF extraction, typical DA methods include channel models such as AWGN~\cite{xie2021generalizable, yu2019multi, yu2019robust} and Gaussian FIR filtering~\cite{shen2022towards, al2021deeplora, soltani2020more}. This type of DA is easy to implement and provides certain performance improvements, but it relies on qualitative prior knowledge, which may cause noticeable information loss. For instance, if there is a mismatch between the channel models adopted in DA and those encountered in the training dataset, some important features, i.e., those robust to real-world channels but sensitive to the adopted channel models  in DA, will be discarded. On the other hand, the features that are robust to the channel model assumed in DA are not necessarily robust to real-world situations. This means that features sensitive to real-world environments can still remain. For these cases, the improvement achieved by DA diminishes, which in turn degrades the RFF authentication.}


\bflag{To address the mismatch between handcrafted DA and the training data while preventing ML-based RFF from overfitting, learning-based DA could be a promising solution based on learning from existing data to generate synthetic data. However existing learning-based DA methods still have limitations. For example, generative models~\cite{kingma2013auto,goodfellow2014generative, chen2016infogan,higgins2016beta} learn to map a low-dimensional latent space to the data space for data generation. Such methods usually suffer from problems of low-quality generation, e.g., blurry image generation~\cite{kingma2013auto} or unstable training and mode collapse in~\cite{goodfellow2014generative}. Alternatively, feature space augmentation~\cite{devries2017dataset} generates augmented data by manipulating the feature vector space rather than the data space, but it is hard to interpret and data-space augmentation achieves better performance~\cite{shorten2019survey}. Another approach is adversarial training~\cite{madry2017towards}, which takes adversarial examples from attacks as augmented data to impose a strong regularization effect on training for the robust model at the cost of performance impairment from the abandonment of the features sensitive to attacks~\cite{tsipras2018robustness}.} 

\bflag{In this paper, we propose a novel framework based on disentangled representation  learning~\cite{higgins2018towards}. The proposed learning framework is tailored for open-set RFF authentication taking advantage of the above learning-based DA methods while avoiding their shortcomings. }

\paragraph{Disentangled representation learning}
\bflag{Disentangled representation~(DR) learning, a combination of representation learning~\cite{bengio2013representation} and generative models~\cite{kingma2013auto,goodfellow2014generative}, projects the observed data onto a lower dimensional form and breaks down or disentangles the data into meaningful underlying factors for subsequent data reconstruction~\cite{liu2021tutorial, higgins2018towards}. A representative example is unsupervised DR, e.g., the generative models in~\cite{kingma2013auto, chen2016infogan, higgins2016beta}, which learn the input-to-latent-variable mapping based on the assumption of a prior distribution over the latent space. Unsupervised DR methods~\cite{kingma2013auto, chen2016infogan, higgins2016beta} have been targeted for applications in wireless communication, including but not limited to indoor localization~\cite{chen2021a}, joint source coding~\cite{choi2019neural, saidutta2021joint}, as well as unsupervised RF fingerprinting~\cite{gong2020unsupervised}. Since the learning of these frameworks is unsupervised, the semantic information contained in each dimension of the latent space is uncontrollable and uncertain. }

\bflag{Another type of DR is self-supervised DR~\cite{xie2019adversarial, denton2017unsupervised, chou2018multi, tran2017disentangled, tran2017disentangled, mitrovic2020representation, chen2022scalable}, which disentangles data into representations with certain controllable meanings by introducing domain-specific or task-specific priors. For example, for video prediction, the DR method in~\cite{xie2019adversarial, denton2017unsupervised} was exploited to disentangle moving objects in a surveillance video from a static background. For voice conversion, the DR method in~\cite{chou2018multi} disentangles the speech signals into speaker identities and speaker-independent representations. The DR method in~\cite{tran2017disentangled} extracted pose information and facial identities from images to synthesize identity-preserving faces and achieve pose-invariant facial recognition.
Indeed, with an appropriate DR design, one can obtain deep models that are robust to representations from unseen domains that conventional DA techniques cannot always achieve~\cite{higgins2018towards}. However, these methods are domain- or task-specific and have various limitations in expanding to other domains. For example, the DR method in~\cite{tran2017disentangled} needs the training data to contain multi-perspective labels for learning disentanglement, which is not easily satisfied in practice.}

\bflag{Since these DR methods are based on generative models and manipulating the feature vector space rather than the data space to generate data, they still suffer from low-quality generation and inefficient regularization when they are used for DA.}

\subsection{Main contributions}
\bflag{In this paper, we adapt the self-supervised DR for open-set RFF extraction.} The main contributions of this work are threefold.

1) We propose to disentangle the received signal into \emph{device-relevant} and \emph{device-irrelevant} representations via three DNNs. 
The device-relevant representation refers to the essential information for effective RFF and the device-irrelevant representation represents the ``background'' of the signal, which contains both noise and the effects of RF propagation. \bflag{Inspired by~\cite{bertran19a}, we first adopt information obfuscation learning to enable device-relevant information suppression and modification in the data space. Moreover, we simplify this learning by reusing the discriminative RFF extractor~\cite{xie2021generalizable} for device-relevant information estimation, such that only three neural networks are necessary to achieve this disentanglement. Since device-irrelevant information is dominant in the received signal, we adopt two domain-preserving networks for device-irrelevant information preservation and high-quality signal generation.}

 
2) We exploit the fact that even though the devices may be located in similar environments or nearby each other, distinctions in the ``background'' of the received signals will still exist. Based on this observation, the proposed DR learning framework shuffles the ``backgrounds'' within the original training data, which implicitly synthesizes more data and maximally enlarges the data space in a data-driven manner. \bflag{Since the proposed framework provides signal diversity from the training data itself, which is more realistic than from handcrafted channel models, it is therefore less destructive to the features that are robust to real-world channels. }

3) We evaluate the proposed methods using a real-world testbed. The experiments verify that the proposed framework outperforms  conventional DL-RFFs for unknown channels. The implicit data augmentation in the proposed DR learning framework can significantly reduce the overfitting of known channels and provide a better trade-off between robustness and performance than the conventional methods.
\begin{table*}[t]  

\caption{\bflag{Notations used throughout the paper}}

\centering
\resizebox{\linewidth}{!}{	
\begin{tabular}{ll|ll}  
\toprule   
Notation & Definition & Notation & Definition  \\  
\midrule
$\mathbf{x}$ & Received signal & $\mathbf{n}$ & Gaussian noise \\
$\mathbf{y}$ & One-hot encoding of the identity of a known device & $D(\cdot, \cdot)$ & Distance function \\
$(\mathbf{x}_i, \mathbf{y}_i)$ & The $i$-th sample from a dataset & $F(\cdot)$ & RFF extractor, with input $\mathbf{x}$ and output $\mathbf{z}$   \\
$\mathbf{z}$ & Radio fingerprint. $\mathbf{z}_i$ denotes the radio fingerprint from $\mathbf{x}_i$  & $Q(\cdot, \mathbf{n})$ & Background extractor, with input $\mathbf{x}$ and output $\overline{\mathbf{x}}$   \\
$\overline{\mathbf{x}}$ & Background signal. $\overline{\mathbf{x}}_i$ denotes the background signal from $\mathbf{x}_i$  &   $G(\cdot, \cdot)$ & Signal generator, with $\mathbf{z}$ and $\overline{\mathbf{x}}$ as inputs and output $\hat{\mathbf{x}}$  \\
$\hat{\mathbf{x}}$ & Synthetic signal. $\hat{\mathbf{x}}_{i,j}$ is synthesized by using $\mathbf{z}_i$ and $\overline{\mathbf{x}}_j$              & $p_{\mathbf{W}}\left(\mathbf{y} | \mathbf{z}\right)$ & Auxiliary linear classifier with learnable parameters $\mathbf{W}$\\
%
\bottomrule  
\end{tabular}
}
\label{tb:notation}
\vspace{-0.5cm}
\end{table*}
The rest of this paper is organized as follows. Section II describes the system model. Section III elaborates the details of the proposed method, and Section VI presents the experimental tests and results. Finally, Section V concludes this paper.

\emph{Notation:} Throughout this paper, boldface lower case letters denote a random column vector, $\veca^{\top}$ and $\| \veca \|$ denote the transpose and the $l_2$-norm of vector $\veca$, $\mathcal{I}(\mathbf{a}; {\mathbf{b}})$ denotes the mutual information between $\mathbf{a}$ and ${\mathbf{b}}$, $\mathcal{N}(\mathbf{0}, \mathbf{I})$ denotes the real-valued normal distribution with zero mean and identity covariance, the operator $[\cdot]_+$ is defined as $[\cdot]_+ \triangleq \max \{\cdot, 0\}$
 and $\nabla_A (\mathcal{L})$ represents the gradient of $\mathcal{L}$ with respect to the trainable parameters of the DNN module A. \bflag{Additional notation is defined in Table~\ref{tb:notation}.}


\section{System Overview}
\begin{figure}[t]
	\centering
	\includegraphics[width=0.95\linewidth]{\rootpath/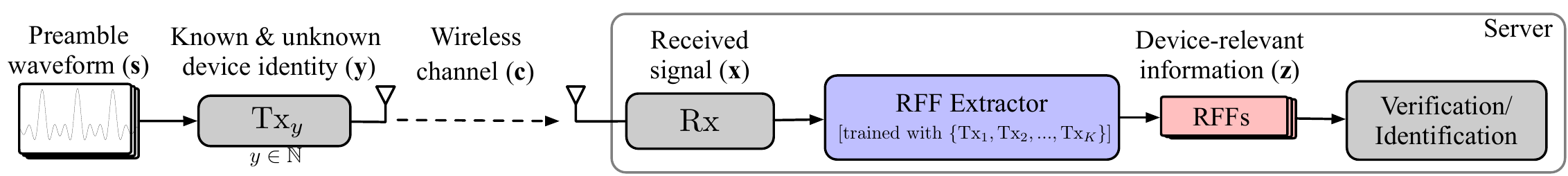}
	\caption{\bflag{The diagram of an open-set RFF authentication system.}}
	\label{fig:sys}
	\vspace{-0.5cm}
\end{figure}

\subsection{Open-set RFF Authentication}
\bflag{We consider an open-set RFF authentication system as depicted in Fig. 1 that consists of a set of transmitting terminals and one server.} Formally, given a preamble of length $M$, denoted by $\mathbf{s}\in \mathbb{C}^M$, the received signals $\mathbf{x} \in \mathbb{C}^M$ can be written as 
\begin{equation}
\begin{aligned}
	\vecx = f_{\vecc}\big( f_{y}(\mathbf{s})\big),
	\label{eq:sys}
\end{aligned} 
\end{equation}
where $f_{\vecc}: \mathbb{C}^M \rightarrow \mathbb{C}^M$ is the functional representation of the wireless channel\footnote{\bflag{Note that $f_{\vecc}$ can be an AWGN channel or a general multipath channel. However, in real-world situations, the wireless channel can be time-varying and more complicated, which is challenging to model accurately.}} and $f_{y}:\mathbb{C}^M \rightarrow \mathbb{C}^M$ represents the effects imposed by the hardware characteristics of the transmitter. 
The authentication system uses the RFF extractor to separate the inherent hardware characteristics from the received signal $\mathbf{x}$, i.e., the RFF. Mathematically, we denote this by
\begin{equation} 
\begin{aligned}
	\vecz = F(\vecx),
	\label{eq:extractor}
\end{aligned} 
\end{equation}
where $F: \mathbb{C}^M \rightarrow \mathbb{R}^d$ is the RFF extractor implemented using some type of DNN, which \bflag{is trained by using $K$ known devices, i.e., $\{\text{Tx}_1, \text{Tx}_2,..., \text{Tx}_K\}$, where $y \in  \mathbb{N}$ indicates the device identity and $\mathbf{Tx}_y$ denotes the $y$-th transmitter.} The obtained RFF $\vecz$ of length $d$ is then compared against known RFFs using some distance function in the final step of the device authentication process. In particular, given a distance function \bflag{$D(\cdot; \cdot)$}, verification of RFF $\vecz_i$ against RFF $\vecz_j$ can be achieved as follows,
\begin{equation}
\begin{cases}
D\left(\vecz_i; \vecz_j \right) \leq T \quad \Rightarrow \quad {\vecz_i\text{ and }\vecz_j\text{ from the same device}}, \\
D\left(\vecz_i; \vecz_j \right) > T  \quad \Rightarrow  \quad {\vecz_i\text{ and }\vecz_j\text{ from different devices}}, \\
\end{cases}
\label{eq:verification}
\end{equation}
where $T$ is a threshold that is to be optimized based on the given training data. 
To achieve satisfactory authentication performance, the RFF $\vecz$ should not only be sufficiently discriminative, but the value of $\vecz$ should only depend on the hardware properties encoded by $f_{y}(\cdot)$. In other words, this requires the RFF extractor $F(\cdot)$ to maximally mitigate the impact of the wireless channels, i.e., $f_\vecc(\cdot)$, while retaining the unique characteristics of the device hardware. 
\subsection{ML RFF Extractor}
In order to retain device-relevant information and to obtain discriminative RFFs, a maximum likelihood~(ML) RFF extractor was previously proposed in \cite{xie2021generalizable}. \bflag{Consider a training set $\mathcal{T}=\{(\vecx_{i}, \vecy_{i})\}_{i=1}^N$ with $N$ samples from $K$ known terminals, where $\vecy_i \in \{\mathbf{e}_{y} : y=1,....,K\}$, and $\mathbf{e}_{y}$ is a vector with an ``1'' in position $y$ and zeros elsewhere, indicating which of the $K$ known terminals the signal corresponds to.} The ML RFF extractor $F(\cdot)$ in \cite{xie2021generalizable} is obtained by solving the optimization problem:
\begin{equation}
\begin{aligned}
	\max_{F, \mathbf{W}} \quad \frac{1}{N} \sum^{N}_{i=1} \ln p_\mathbf{W}(\vecy_{i}|\vecz_{i}), \quad \text{s.t.} \quad \vecz_{i} = F(\vecx_{i}),
	\label{eq:obj}
\end{aligned} 
\end{equation}
where \bflag{the conditional probability $p_\mathbf{W}(\vecy|\vecz)$ can be viewed as an auxiliary classifier\footnote{\bflag{This conditional probability can be rewritten as the likelihood $p_{\Theta}(\mathbf{y}|\mathbf{x})$, where $\Theta=\{F, \mathbf{W}\}$. In this sense, optimizing the loss $-\sum_{i=1}^N \ln p_\mathbf{W}(\mathbf{y}_i|\mathbf{z}_i)$ as in (4) corresponds to the ML estimation of the parameters $F$ and $\mathbf{W}$.}}} that establishes the relationship between RFF $\vecz$, and the device identity $\vecy$.  $\mathbf{W}$ is a set of trainable parameters in the auxiliary classifier. 

\textbf{\emph{Hypersphere projection}}: To achieve device discrimination with $\vecz$, $p_\mathbf{W}(\vecy|\vecz)$ is implemented in the form of a softmax probability~\cite{ranjan2017l2}:
\begin{equation}
\begin{aligned}
	p_\mathbf{W}(\bflag{y}|\vecz) = \frac{\exp \left\{\overline{\mathbf{w}}_{y}^{\top} \overline{\mathbf{z}}\right\}}{\sum_{j} \exp \left\{\overline{\mathbf{w}}_{j}^{\top} \overline{\mathbf{z}}\right\}}, \quad \forall y=1,2,...,K,
\end{aligned} 
\end{equation}
where we define
\begin{equation}
\begin{aligned}
\overline{\mathbf{w}}=\frac{\mathbf{w}}{\|\mathbf{w}\|} \quad \text{and}\quad \overline{\mathbf{z}}= \delta \frac{\mathbf{z}}{\|\mathbf{z}\|},
\end{aligned}
\label{eq:normed}	
\end{equation}
and $\delta > 0$ is a hyper-parameter that controls the norm of $\overline{\mathbf{z}}$. Here, $\mathbf{W}=\{\{\mathbf{w}_{j}\}_{j=1}^{K}\}$ represents the parameters of the softmax classifier. The normalization in (\ref{eq:normed}) is also known as hypersphere projection~(HP), where $\delta$ is the radius of the hypersphere. Note that HP is popularly adopted in facial recognition~\cite{ranjan2017l2, deng2019arcface, liu2017sphereface}, and \bflag{it regulates the norms of the feature vector to guarantee that (\ref{eq:obj}) is equivalent to using the cosine distance,  i.e., $D\left(\vecz_i; \vecz_j \right)=1-\frac{\vecz_i^{\top} \vecz_j}{\|\mathbf{z}_i\|\|\mathbf{z}_j\|}$ in (\ref{eq:verification}), to perform discrimination of the RFFs.} Using this formulation, the RFF extractor $F(\cdot)$ can maximally retain the device-relevant information in $\vecx$ to improve the quality of the RFF discrimination.
 
Given sufficient training data representative of the entire data space of channel realizations, the ML RFF extractor is the best in the sense of probability of successful RFF discrimination~\cite{barber2012bayesian}. However,  collecting sufficient data to capture the entire dynamic channel space in real-world scenarios is expensive and impractical, especially for massive IoT applications. If the training data is insufficiently rich, e.g., if it is collected only from simple LoS-dominated channels, the ML RFF extractor will tend to overfit this non-representative channel statistic existing in the training data.
More importantly, generalizations of this approach to other types of channels, e.g., dispersive multipath channels, is limited.

\section{Disentangled Representation Learning for RFF Extraction}
In this section, we propose a DR learning framework to improve the generalizability of DL-RFFs adapted to practical wireless channels drawn from distributions that are unavailable or unseen in the training data. We first introduce the main idea of the design and then elaborate on the details of the proposed framework.
\subsection{Proposed DR Learning Framework}

The proposed DR learning framework first learns to factor the received signal into two disjoint parts, i.e., a \emph{device-relevant} representation and a \emph{device-irrelevant} representation, and then to synthesize augmented training signals given these representations. Here, the \emph{device-relevant} representations are the RFFs and the \emph{device-irrelevant} representations are regarded as other ``background'' information associated with the received signals such as that associated with the propagation environment.  This factorization allows us to swap out the backgrounds of different signals and thus create new data for augmenting the training set.

In practice, due to slight differences in the angle and position of the device antennas when acquiring signals, even if the training data from all the devices are collected in a simple LoS scenario, distinctions between their channels can still exist. Thus, the training dataset may still contain multiple different backgrounds among its various signals.
By disentangling the signals into device identities and backgrounds, we can generate augmented signals that preserve device identity and that are representative of data that would be generated by the device under every possible background in the training dataset. Since the background distinctions are essentially distinction in the channels, the RFF extractors trained by these augmented signals are encouraged to ignore these channel distinctions and extract channel-invariant features based solely on the RFFs
A promising observation from our experiments in Section IV is that the channel variations under the LoS assumption are sufficiently rich to improve the generalizability of the RFF extractor in the test sets. 
\begin{figure*}[t]
	\centering
	\includegraphics[width=0.90\linewidth]{\rootpath/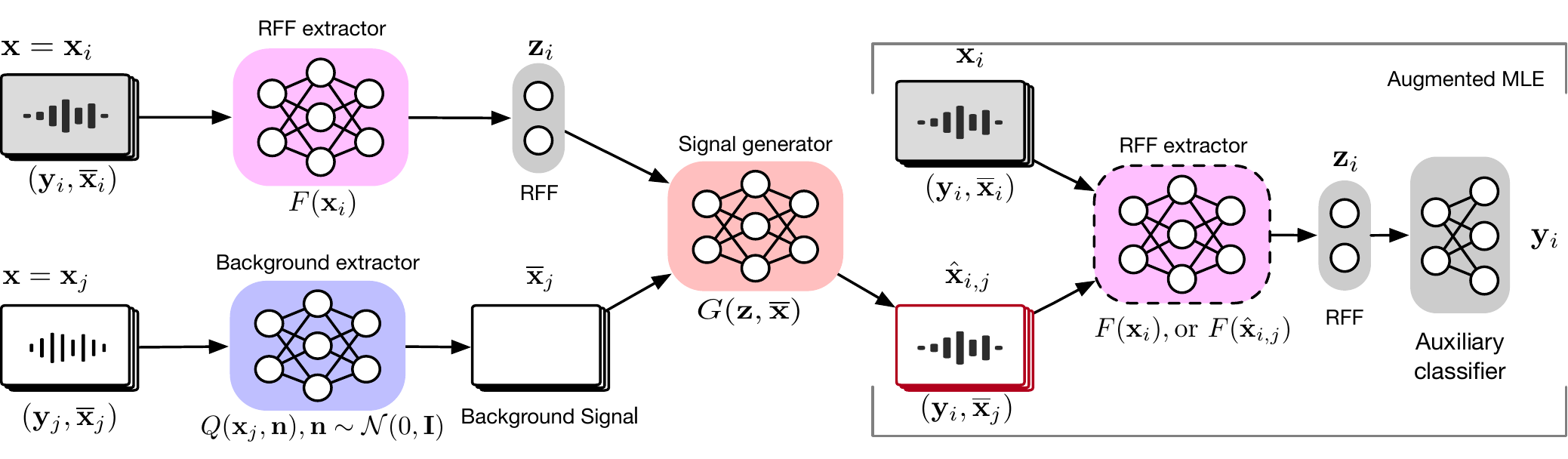}
	\caption{\bflag{The proposed DR learning framework for RFF extraction~({\bf F-step}). Given the two received training signals, the RFF and the background signal are extracted by the RFF extractor~(pink) and the background extractor~(blue), respectively. A synthetic signal is generated by feeding the RFF and the background signal to the signal generator~(red). The raw and synthetic signals, which have the same RFF but different signal backgrounds, are used to train the RFF extractor~(pink dotted box).  }}
	\label{fig:diagram}
	\vspace{-0.5cm}
\end{figure*}
The proposed framework, as depicted in Fig. \ref{fig:diagram}, consists of three main DNN modules, i.e., $F(\cdot)$, $Q(\cdot, \vecn)$, and $G(\cdot, \cdot)$. We articulate these three modules below.

\emph{a) \textbf{RFF extractor $F(\cdot)$}}: This module, represented by the pink boxes in Fig.~\ref{fig:diagram}, takes signal $\vecx$ as the input and outputs the corresponding RFF $\vecz$ in (\ref{eq:extractor}). The vector $\vecz$ is taken to be the device-relevant information within $\vecx$. \bflag{Besides its usage as an RFF extractor}, this module is also adopted as an adversarial discriminator for estimating how much of the device-relevant information is contained in the background signal from $Q(\cdot, \vecn)$, which is introduced next.

\emph{b) \textbf{Background extractor $Q(\cdot, \vecn)$}}: This module,  shown as the blue box in Fig.~\ref{fig:diagram}, realizes a stochastic mapping which is used for preserving device-irrelevant information while ruling out device-relevant information as much as possible. Given the input signal $\vecx$, the background signal, denoted by $\overline{\vecx}$, is obtained as 
\begin{equation}
\begin{aligned}
\label{eq:rep}
\overline{\vecx}\sim p_Q(\overline{\vecx}| \vecx) \Longleftrightarrow \overline{\vecx} = Q(\vecx, \vecn),\quad \vecn \sim \mathcal{N}(\mathbf{0}, \mathbf{I}).
\end{aligned} 
\end{equation}
This stochastic mapping is also used for sensitive information obfuscation in \cite{bertran19a}. Similarly, the randomness in (\ref{eq:rep}) is introduced for purposefully obfuscating the device-relevant information of $\vecx$. The background signal $\overline{\vecx}$, complementary to the RFF in forming $\vecx$, contains only the device-irrelevant information within $\vecx$, which can capture the joint effects of the wireless channel, noise, the preamble waveform, etc. 

\emph{c) \textbf{Signal generator $G(\cdot, \cdot)$}}: This module, the center red box in Fig.~\ref{fig:diagram}, is adopted for signal reconstruction and generation. The input to this module includes both the RFF and the background signal. Given these mutually complementary representations, the synthetic signal, denoted by $\hat{\vecx}$, is generated by
\begin{equation}
\begin{aligned}
\hat{\vecx}= G(\vecz, \overline{\vecx}).
\end{aligned} 
\end{equation}

With these three modules, the proposed DR learning framework establishes a \bflag{flexible and  convenient} approach to generate augmented signals for improving the robustness of the RFF extraction. Ideally, exponentially more augmented data can be arbitrarily generated from the raw training set by arbitrarily swapping their background signals and introducing randomness. The training in our framework is performed in an iterative manner by the modules. Moreover, by applying the proposed framework, the augmented signals are also dynamically improved during the learning process. Details on the learning algorithm will be introduced later in Section III.E.

Note that the RFF extractor trained within the proposed framework is forced to extract the background-irrelevant~(i.e., device-relevant) RFF information from the signals \bflag{in the training data} and therefore improve its generalizability and robustness. In the following, we refer to this RFF extractor as the \emph{DR-RFF extractor}.

From the above, we see that $\overline{\vecx}$ and $\hat{\vecx}$ can both be adopted to extract the RFFs via $F(\cdot)$. Thus, in order to preserve the inference capability of $F(\cdot)$ for $\overline{\vecx}$ and $\hat{\vecx}$, we must restrict $G(\cdot, \cdot)$ and $Q(\cdot, \vecn)$ to be domain-preserving~\cite{bertran19a}, e.g., signal-to-signal transformations.
The design of the learning procedure and the detailed structures of each of the three modules, i.e., $F(\cdot)$, $Q(\cdot,\vecn)$, and $G(\cdot, \cdot)$, are elaborated in the following. 
\subsection{Learning DR-RFF Extractor $F(\cdot)$}
Given the raw data pair in the training set, $(\vecx, \vecy)\in \mathcal{T}$, the augmented signals, $\hat{\vecx}$, are generated by the proposed DR learning framework, and $\hat{\vecx}$ contains the same device identity information but a different signal background compared to $\vecx$. The goal of the proposed DR-RFF extractor is to distill the same device-relevant information from both $\vecx$ and $\hat{\vecx}$ while mitigating the impact of their backgrounds. From the perspective of information theory, this goal can be achieved by maximizing the mutual information~\cite{mackay2003information} between the corresponding RFFs and the device identity $\vecy$, as follows:
\ifthenelse{\boolean{isdouble}}{
\begin{equation}
\label{eq:obj_F}
\begin{aligned}
\max_{F}& \quad \lambda \mathcal{I}(\mathbf{y};\vecz) + (1-\lambda) \mathcal{I}( \mathbf{y}; \hat{\vecz}) \\
\text{s.t.}& \quad \vecz = F(\mathbf{x}), \quad \hat{\vecz} = F(\hat{\mathbf{x}}),
\end{aligned} 
\end{equation}
}{
\begin{equation}
\label{eq:obj_F}
\begin{aligned}
\max_{F} \quad \lambda \mathcal{I}(\mathbf{y};\vecz) + (1-\lambda) \mathcal{I}( \mathbf{y}; \hat{\vecz}) \quad
\quad \text{s.t.} \quad \vecz = F(\mathbf{x}), \quad \hat{\vecz} = F(\hat{\mathbf{x}}),
\end{aligned} 
\end{equation}
}
where $0\le \lambda < 1$ is a hyper-parameter that balances the learning effects for the raw and augmented signals. The first term in the objective function of (\ref{eq:obj_F}), measuring the amount of device-relevant information extracted from the raw signal, is the same RFF learning objective as the one in our previous work~\cite{xie2021generalizable}. The second term is the objective corresponding to the proposed augmented training. It encourages the RFF extractor $F(\cdot)$ to extract the same identity from the raw and augmented signals, which is the key to avoid overfitting of $F(\cdot)$ to the specific channel statistics embedded in the raw data. 

To facilitate the applications of DNNs to solve the problem in (\ref{eq:obj_F}), we now reformulate it to obtain a tractable data-driven objective function. Mathematically, as exemplified in Fig.~\ref{fig:diagram}, we draw two arbitrary signals from devices $\vecy_i$ and $\vecy_j$, collected under different propagation environments in the training dataset, i.e.,
\begin{equation}
\begin{aligned}
	(\vecx_i, \vecy_i)\in \mathcal{T},\quad (\vecx_j, \vecy_j) \in \mathcal{T}.
\end{aligned} 
\end{equation}
From the left hand side of Fig.~\ref{fig:diagram}, the device-relevant RFF and the device-irrelevant background representations are respectively extracted as
\ifthenelse{\boolean{isdouble}}{
\begin{align}
	\vecz_i &= F(\vecx_i),\\
	\overline{\vecx}_j &= Q(\vecx_j,\mathbf{n}),\text{ for } \mathbf{n} \sim \mathcal{N}(\mathbf{0}, \mathbf{I}).
\end{align}
}{
\begin{align}
	\vecz_i = F(\vecx_i), \quad \overline{\vecx}_j &= Q(\vecx_j,\mathbf{n}),\text{ for } \mathbf{n} \sim \mathcal{N}(\mathbf{0}, \mathbf{I}).
\end{align}
}
The synthetic signal, $\hat{\vecx}_{i,j}$, is generated from the above two representations via the signal generator as
\begin{equation}
\begin{aligned}
	\hat{\vecx}_{i,j} = G(\vecz_i,\overline{\vecx}_j),
\end{aligned} 
\end{equation}
where $\hat{\vecx}_{i,j}$ represents a received signal that is transmitted by device $\vecy_i$ but undergoes the same propagation channel as device $\vecy_j$. In principle, the module $G(\cdot, \cdot)$ learns to mimic a transmission from device $\vecy_i$ under the propagation environment of another device $\vecy_j$. 

Following the derivations in \cite{alemi2017deep} and \cite{xie2021generalizable}, we reformulate (\ref{eq:obj_F}) into an ML estimation problem as in (\ref{eq:obj}). The learning problem for $F(\cdot)$ in (\ref{eq:obj_F}), denoted by $\mathcal{L}_F$, is rewritten as 
\ifthenelse{\boolean{isdouble}}{
\begin{equation}
\label{eq:pbl_F}
\begin{aligned}
	 \mathcal{L}_F \triangleq  \frac{1}{N^2} \sum_{i=1}^{N} \sum_{j=1}^{N} \bigg[&\lambda \log  p_\W (\vecy_i|F(\mathbf{x}_i)) \\
	 &+ (1-\lambda) \log p_\W(\vecy_i|F(\hat{\mathbf{x}}_{i,j}))\bigg].
\end{aligned} 
\end{equation}
}{
\begin{equation}
\label{eq:pbl_F}
\begin{aligned}
	 \mathcal{L}_F \triangleq  \frac{1}{N^2} \sum_{i=1}^{N} \sum_{j=1}^{N} \bigg[&\lambda \ln  p_\W (\vecy_i|F(\mathbf{x}_i)) + (1-\lambda) \ln p_\W(\vecy_i|F(\hat{\mathbf{x}}_{i,j}))\bigg].
\end{aligned} 
\end{equation}
}
It is proved in \cite{xie2021generalizable} that $-\mathcal{L}_F$ is essentially a variational lower bound of (\ref{eq:obj_F}). 
Learning this objective is equivalent to optimizing the original problem in (\ref{eq:obj_F}) while circumventing the intractable computation in (\ref{eq:obj_F}). The value of $\mathcal{L}_F$ can be easily calculated using data samples. Thus, we can optimize the DR-RFF extractor $F(\cdot)$ using (\ref{eq:pbl_F}) and the gradient descent algorithm, e.g., Adam~\cite{kingma2014adam}, with the training data set. 
Note that we detach every augmented signal $\hat{\vecx}_{i,j}$ from the generation process and treat it as an independent sample in training $F(\cdot)$. Therefore, we do not backtrack to the representation extraction when the back-propagation goes through the computational graph during the training of $F(\cdot)$.
\begin{figure*}[t]
	\centering
	\includegraphics[width=0.98\linewidth]{\rootpath/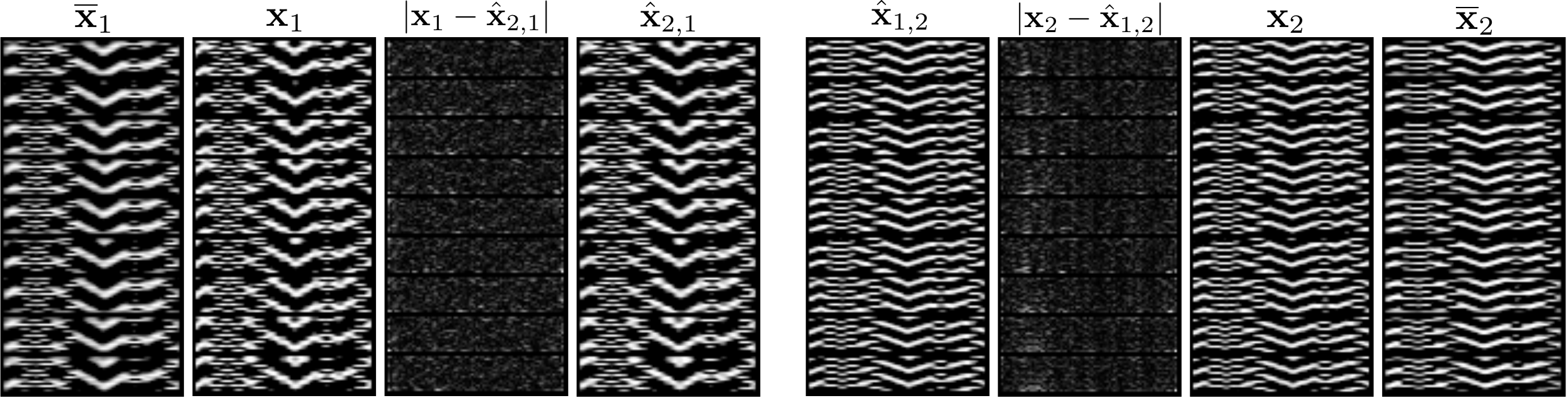}
	\caption{Visualization of raw signals~($\vecx_{1}$ and $\vecx_{2}$), background signals~($\overline{\vecx}_{1}$ and $\overline{\vecx}_{2}$), and synthetic signals~($\hat{\vecx}_{1,2}$ and $\hat{\vecx}_{2,1}$).}
	\label{fig:vis}
	\vspace{-0.5cm}
\end{figure*}

One additional trick for the design of $F(\cdot)$ is that the HP operation in (6) applied to $p_\W (\vecy|\vecz)$ is indispensable. For a successful disentanglement, the RFFs extracted by $F(\cdot)$ should contain as little of the background information as possible. This means that $\vecz_i$ and $\vecz_{i,j}$ should be close to each other in terms of the cosine distance adopted in (\ref{eq:verification}). The HP operation is necessary for obtaining discriminative RFFs, which is the key for aggregating the RFFs from the same device~(e.g., $\vecz_i$ and $\vecz_{i,j}$) under the cosine distance.

To more intuitively explain the intrinsic mechanisms of the proposed DR learning framework, we visualize the real part of the raw signals, background signals, and the synthetic signals in Fig. \ref{fig:vis}. Comparing the raw signals with the background signals, we find that the textures of the signal backgrounds are dominant in the augmented signals. The difference signals, i.e., $|\vecx_{1} - \hat{\vecx}_{1, 2}|$ and $|\vecx_{2} - \hat{\vecx}_{2, 1}|$ in Fig. \ref{fig:vis}, reveal the embedded RFFs in the augmented signals and indicate that the device-relevant information in the signals is imperceptible.

\subsection{Learning Background Extractor $Q(\cdot, \vecn)$}
The goal of $Q(\cdot, \vecn)$ is to extract the background signals $\overline{\vecx}$ from the input signals $\vecx$. The background signal $\overline{\vecx}$ is expected to preserve as much information as possible from the inputs after removing the device-relevant information. Mathematically, this goal can be formulated as 
\begin{equation}
\begin{aligned}
\label{eq:bg}
\max_{Q} \quad \mathcal{I}(\mathbf{x} ; \overline{\mathbf{x}}),\quad \text{s.t.} \quad \mathcal{I}(\mathbf{y} ; \overline{\mathbf{x}})<\epsilon,
\end{aligned} 
\end{equation}
where $\mathcal{I}(\mathbf{x} ; \overline{\mathbf{x}})$ and $\mathcal{I}(\mathbf{y} ; \overline{\mathbf{x}})$ respectively quantify the amount of information that the background signal $\overline{\vecx}$ contains about that the original signal $\vecx$ and the identity $\vecy$, and $\epsilon\ge 0$ is a hyper-parameter that controls the amount of device-relevant information that remains in $\overline{\vecx}$. To facilitate the subsequent development, we further relax the problem in (\ref{eq:bg}) and convert it into an unconstrained problem by using a quadratic penalty~\cite{boyd2004convex} as follows: 
\begin{equation}
\label{eq:bg2}
\begin{aligned}
\max_{Q} \quad \mathcal{I}(\mathbf{x} ; \overline{\mathbf{x}}) - \alpha \big[\mathcal{I}(\mathbf{y}; \overline{\mathbf{x}})-\epsilon\big]_{+}^2,
\end{aligned} 
\end{equation}
where $\alpha>0$ is the penalty parameter. The problem in (\ref{eq:bg2}) is equivalent to the original problem in (\ref{eq:bg}) when $\alpha \rightarrow \infty$. 
This formulation is connected with the information bottleneck~(IB) approach~\cite{tishby2000information}, which was initially designed for random variable compression and has been exploited for exploring the intrinsic learning mechanism of DNNs~\cite{tishby2015deep}, for training robust DNNs~\cite{alemi2017deep}, \bflag{and for sensitive information obfuscation~\cite{bertran19a}}. It is typically used for finding the best trade-off between model accuracy and representation complexity. In (\ref{eq:bg}), we exploit this IB-like formulation to strike a balance between the ``signal reconstruction quality''~(i.e., the maximization of $\mathcal{I}(\mathbf{x} ; \overline{\mathbf{x}})$) and the  ``elimination of device-relevant information''~(i.e., the minimization of the penalty term) to achieve the disentanglement.
\ifthenelse{\boolean{isdouble}}{}{
\begin{figure*}[t]
	\centering
	\includegraphics[width=0.9\linewidth]{\rootpath/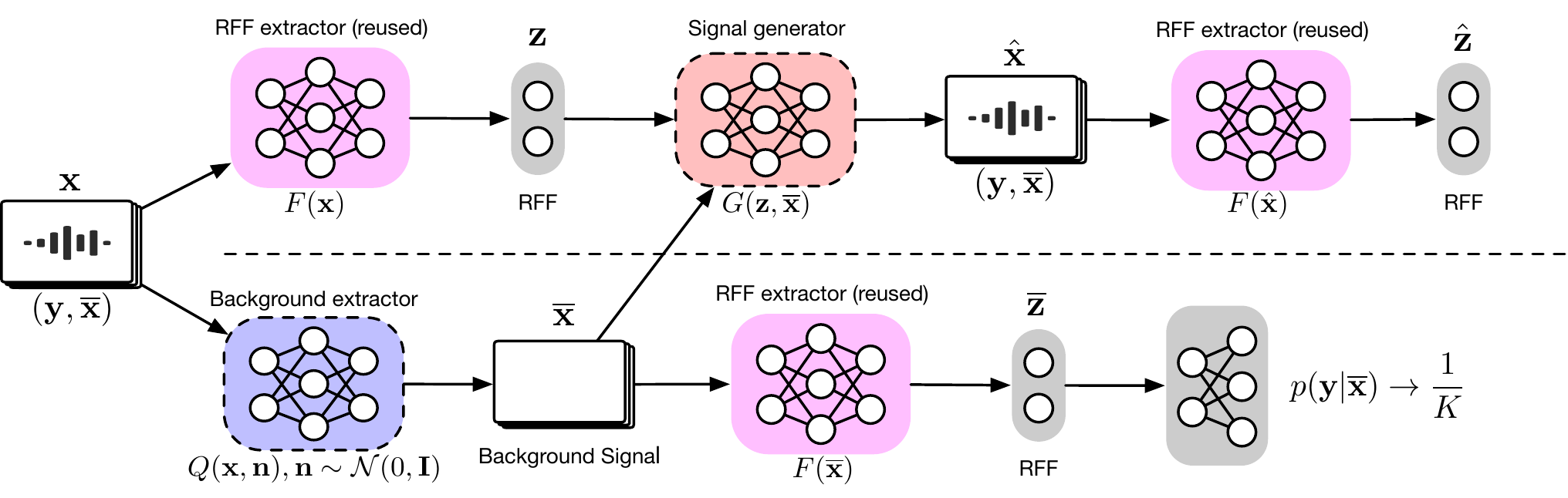}
	\caption{\bflag{The proposed DR learning framework for RFF extraction~({\bf Q/G-step}). Given a received signal, the background extractor~(blue dotted box) learns to extract the background signal that cannot provide any discriminative ability to the fixed RFF extractor~(pink). The signal generator~(red dotted box) learns to reconstruct the signal using the given RFF and the background signal. The reconstructed signal should also preserve the same RFF as the original signal. }}
	\label{fig:diagram2}
	\vspace{-0.5cm}
\end{figure*}
}
In order to facilitate the model training, we need to reformulate (\ref{eq:bg2}) by rewriting the learning objective with respect to only the training data. We rewrite the two terms in (\ref{eq:bg2}) into a data-driven form by respectively applying the techniques of \emph{information maximization}~\cite{agakov2004algorithm} and \emph{adversarial learning}~\cite{goodfellow2014generative, donahue2016adversarial} as discussed in the following.
\ifthenelse{\boolean{isdouble}}{\begin{figure*}[t]
	\centering
	\includegraphics[width=0.9\linewidth]{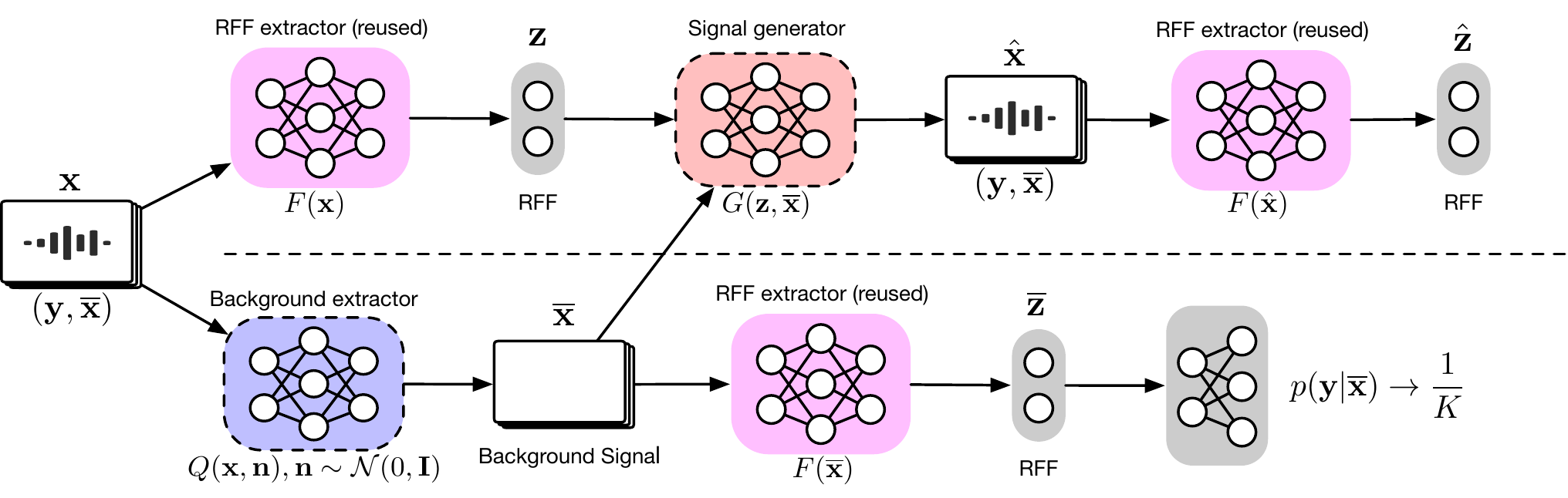}
	\caption{\bflag{The proposed DR learning framework for RFF extraction~({\bf Q/G-step}). Given a received signal, the background extractor~(blue dotted box) learns to extract the background signal that cannot provide any discriminative ability to the fixed RFF extractor~(pink). The signal generator~(red dotted box) learns to reconstruct the signal using the given RFF and the background signal. The reconstructed signal should also preserve the same RFF as the original signal. }}
	\label{fig:diagram2}
	\vspace{-0.5cm}
\end{figure*}}{}
\paragraph{\textbf{Information maximization}} We begin with the calculation of the first term in (\ref{eq:bg2}). Due to the intractable conditional distribution $p(\vecx|\overline{\vecx})$, it is computationally expensive to directly calculate $\mathcal{I}(\mathbf{x} ; \overline{\mathbf{x}})$. One common approach   to address this problem is to adopt a tractable variational distribution $q(\vecx|\overline{\vecx})$ to replace $p(\vecx|\overline{\vecx})$. This replacement yields a tractable variational lower bound for $\mathcal{I}(\mathbf{x} ; \overline{\mathbf{x}})$ that can be used for indirectly maximizing $\mathcal{I}(\mathbf{x} ; \overline{\mathbf{x}})$. Following \cite{agakov2004algorithm}, we adopt the Gaussian distribution $q(\vecx|\overline{\vecx})=\mathcal{N}(\mathbf{x}|\overline{\mathbf{x}}, \mathbf{I})$ to replace $p(\vecx|\overline{\vecx})$. The resultant variational lower bound, denoted by $-\mathcal{L}_\text{v}$, is 
\begin{align}
\label{eq:lb}
\max_{Q} &\quad \mathcal{I}(\mathbf{x} ; \overline{\mathbf{x}})\nonumber \\
=& \max_{Q} \quad \big\{ h(\mathbf{x})- h(\mathbf{x}|\overline{\mathbf{x}})\big\}\nonumber \\
=& \max_{Q} \quad \big\{ h(\mathbf{x}) + \E _{p(\mathbf{x}, \overline{\mathbf{x}})} [\ln p(\mathbf{x}|\overline{\mathbf{x}})]\big\} \nonumber\\
\overset{(a)}{\ge}& \max_{Q} \quad \big\{ h(\mathbf{x}) + \E _{p_Q(\overline{\mathbf{x}}|\vecx)p(\vecx)} [\ln \mathcal{N}(\mathbf{x}|\overline{\mathbf{x}}, \mathbf{I})]\big\} \nonumber\\
\overset{(b)}\propto & \max_{Q} \quad \bigg\{\underbrace{-\underset{{\mathbf{x}}\in \mathcal{T}, \mathbf{n}\sim\mathcal{N}(\mathbf{0}, \mathbf{I})}{\E} \bigg[ \| \mathbf{x}-Q(\mathbf{x}, \mathbf{n}) \|^2 \bigg]}_{-\mathcal{L}_\text{v}} + c \bigg\},
\end{align} 
where $h(\cdot)$ is the differential entropy~\cite{mackay2003information}, $c$ is a constant that can be ignored, $(a)$ follows from the nonnegativity of the Kullback-Leibler divergence~(KLD), i.e.,
\begin{equation}
\label{eq:kl}
\begin{aligned}
\mathcal{D}_{\text{KL}}\big( p(\mathbf{x}|\overline{\mathbf{x}}) \| q(\mathbf{x}|\overline{\mathbf{x}})\big) = \E _{p(\mathbf{x})} \bigg[\ln \frac{p(\mathbf{x}|\overline{\mathbf{x}})}{\mathcal{N}(\mathbf{x}|\overline{\mathbf{x}}, \mathbf{I})}\bigg] \ge 0,
\end{aligned}
\end{equation}
and $(b)$ follows by adopting the re-parameterization in (\ref{eq:rep}) and dropping the constant terms that are irrelevant to $Q(\cdot, \vecn)$. Therefore, the first term of (\ref{eq:bg2}) can be maximized by minimizing $\mathcal{L}_\text{v}$. 
With this new learning objective $\mathcal{L}_\text{v}$, the first term in (\ref{eq:bg2}) is simplified to a mean-squared error~(MSE) loss in (\ref{eq:lb}), and hence the computational complexity of the optimization is greatly reduced. 


\paragraph{\textbf{Adversarial learning for the penalty}}  Similar to the first term, direct computation of the penalty $\big[\mathcal{I}(\mathbf{y}; \overline{\mathbf{x}})-\epsilon\big]_{+}^2$ is intractable. The function of this term is to suppress any device-relevant information. A variational approach like that used for $I(\vecx;\overline{\vecx})$ in (\ref{eq:lb}) is not effective here since the MSE is not sensitive to the small differences in the device RFFs. Thus, we adopt the adversarial learning technique~\cite{goodfellow2014generative} to calculate this term. More concretely, as depicted in the lower half of Fig. \ref{fig:diagram2}, we reuse the DNN classifier, i.e., $p_\mathbf{W} (\mathbf{y}|F(\cdot))$ in Section III.B, as a discriminator to estimate the posterior $p(\mathbf{y}|\overline{\mathbf{x}})$, as follows:
%
\begin{equation}
\label{eq:lp}
\begin{aligned}
\mathcal{I}(\mathbf{y}; \overline{\mathbf{x}})& = \E_{p(\overline{\mathbf{x}})}\big[ \mathcal{D}_{\text{KL}}\big( p(\mathbf{y}|\overline{\mathbf{x}}) \| p(\mathbf{y})\big)\big]\\
&\overset{(a)}{\approx} \E_{p(\overline{\mathbf{x}})}\big[ \mathcal{D}_{\text{KL}}\big( p_\W (\mathbf{y}|F(\overline{\mathbf{x}})) \| p(\mathbf{y})\big)\big]\\
&\overset{(b)}{=} \underset{ \overline{\mathbf{x}}\sim p_Q(\overline{\mathbf{x}}|\mathbf{x}), (\vecx,\vecy)\in \mathcal{T}}{\E} \bigg[ \ln \frac{p_\W (\mathbf{y}|F(\overline{\mathbf{x}}))}{p(\mathbf{y})} \bigg], 
\end{aligned} 
\end{equation}
where $(a)$ results from using the parameterized conditional distribution $p_\mathbf{W} (\mathbf{y}|F(\overline{\mathbf{x}}))$ to replace the original $p(\mathbf{y}|\overline{\mathbf{x}})$, and $(b)$ follows from using the re-parameterization in (\ref{eq:rep}), i.e., $\overline{\mathbf{x}} = Q(\vecx, \vecn) \text{ for } \vecn \sim \mathcal{N}(\mathbf{0}, \mathbf{I})$. Here, the prior distribution of the identity $\vecy$ can be taken to be a discrete uniform distribution, i.e., {$p(\mathbf{y}=\vecy_{(i)})=\frac{1}{K}, \forall i=1,...,K.$}
Now, we can rewrite the penalty term in (\ref{eq:bg2}) in a data-driven form, denoted by $\mathcal{L}_\text{p}$, as follows
\begin{equation}
\label{eq:lpd}
\begin{aligned}
\mathcal{L}_\text{p} &\triangleq -\bigg[\underset{ \overline{\mathbf{x}}\sim p_Q(\overline{\mathbf{x}}|\mathbf{x}), (\vecx,\vecy)\in \mathcal{T}}{\E} \bigg[ \ln \frac{p_\W (\mathbf{y}|F(\overline{\mathbf{x}}))}{1/K} \bigg] -\epsilon \bigg]_{+}^2.
\end{aligned} 
\end{equation}

Substituting (\ref{eq:lb}) and (\ref{eq:lpd}) into (\ref{eq:bg2}), the learning objective of $Q(\cdot, \vecn)$, denoted by $\mathcal{L}_Q$, is defined as
\begin{equation}
\begin{aligned}
\label{eq:pbl_Q}
\mathcal{L}_Q \triangleq \mathcal{L}_\text{v} + \alpha \mathcal{L}_\text{p}.
\end{aligned} 
\end{equation}
Note that the value of $\mathcal{L}_\text{p}$ depends on the RFF extractor $F(\cdot)$. In this sense, the learning of $Q(\cdot, \vecn)$ can be treated as an adversarial game with two players: $Q(\cdot, \vecn)$ tries to generate the signal to confuse $F(\cdot)$, while $F(\cdot)$, as the adversarial counterpart of $Q(\cdot, \vecn)$, learns to discriminate the signals that are partially generated from $Q(\cdot, \vecn)$.
\subsection{Learning Signal Generator $G(\cdot, \cdot)$}
The only remaining task is the development of the signal generator $G(\cdot, \cdot)$. As depicted in the upper half of Fig.~\ref{fig:diagram2}, the module $G(\cdot, \cdot)$ takes a background signal, $\overline{\mathbf{x}}$, and the corresponding RFF, $\mathbf{z}$, as inputs for reconstructing the raw signal $\mathbf{x}$. The learning problem is designed as follows
\begin{equation}
\label{eq:pbl2}
\begin{aligned}
\max_{G} \quad &\mathcal{I}(\mathbf{x} ; \hat{\mathbf{x}}) + \beta \mathcal{I}(F(\mathbf{x}) ; F(\hat{\mathbf{x}})),
\end{aligned} 
\end{equation}
where $\hat{\mathbf{x}} = G(\mathbf{z}, \overline{\mathbf{x}})$, $\vecz=F(\vecx)$, $\overline{\mathbf{x}}$ is drawn according to $p_{Q}(\overline{\mathbf{x}}|\mathbf{x})$, and $\beta>0$ is a hyper-parameter that balances the two mutual information terms. 
In particular, the maximization of $\mathcal{I}(\mathbf{x} ; \hat{\mathbf{x}}) $ acts to minimize the signal reconstruction loss, which ensures the quality of the synthetic signal. The maximization of $ \mathcal{I}(F(\mathbf{x}) ; F(\hat{\mathbf{x}}))$ ensures that the device-relevant information, i.e., the RFF, is successfully embedded in the synthetic signal. 

Similar to the reformulation of (\ref{eq:lb}), we adopt a variational approximation to solve this problem. In other words, we replace the intractable conditional distributions of the terms in (\ref{eq:pbl2}) with Gaussian distributions and ignore the terms that are irrelevant to $G(\cdot, \cdot)$. This leads to the following data-driven learning objective for $G(\cdot, \cdot)$, denoted by $\mathcal{L}_G$:
\begin{equation}
\label{eq:pbl_G}
\begin{aligned}
&\mathcal{L}_{G} \triangleq \underset{{\overline{\mathbf{\vecx}}\sim p_Q( \overline{\mathbf{\vecx}}|\mathbf{\vecx}), \mathbf{x}}\in \mathcal{T}}{\E} \bigg[ \| \mathbf{x}-\hat{\mathbf{x}} \|^2  + \beta  \| F(\mathbf{x})-F(\hat{\mathbf{x}}) \|^2 \bigg ].
\end{aligned} 
\end{equation}
\subsection{Learning Algorithm}

We now elaborate on the design of the learning algorithm for the proposed DR learning framework. 
In the formulation of the problem proposed thus far, the learning objectives for $G(\cdot, \cdot)$ and $Q(\cdot, \vecn)$ are not mutually exclusive. Given the RFF $\vecz$, improving the quality of the signal reconstruction requires that the other input to $G(\cdot, \cdot)$, i.e., the signal background $\overline{\vecx}$, contains as much information from the original signal as possible. This is also a part of the learning objective of $Q(\cdot, \vecn)$, i.e., $\mathcal{L}_\text{v}$ in (\ref{eq:pbl_Q}). Moreover, driven by the experimental results, we find that jointly training $G(\cdot, \cdot)$ and $Q(\cdot, \vecn)$ can provide less signal reconstruction error and hence higher quality synthesized signals. Based on the above considerations, we merge the learning of $G(\cdot, \cdot)$ and $Q(\cdot, \vecn)$ into one step, referred to as the Q/G-step. 
 
On the other hand, the learning of $F(\cdot)$ requires only the raw signals and the corresponding augmented signals generated by $Q(\cdot, \vecn)$ and $G(\cdot, \cdot)$. Additionally, as a discriminator, $F(\cdot)$ should be made independent of the others. We therefore implement the learning of $F(\cdot)$ in a single step, referred to as the F-step.

In summary, the learning algorithm of the proposed DR learning framework is composed of the following two steps.

\textbf{Q/G-step}: Fixing $F(\cdot)$, we optimize $Q(\cdot, \vecn)$ and $G(\cdot, \cdot)$ to learn to factorize and reconstruct the received signals in the training data set, as depicted in Fig. \ref{fig:diagram2}. By applying the gradient descent algorithm, $Q(\cdot, \vecn)$ and $G(\cdot, \cdot)$ are updated as follows
\ifthenelse{\boolean{isdouble}}{
\begin{equation}
\label{eq:QGstep}
\begin{aligned}
Q \leftarrow Q - \eta \nabla_Q (\mathcal{L}_{Q} + \mathcal{L}_{G}),\\ G \leftarrow G - \eta \nabla_G (\mathcal{L}_{Q} + \mathcal{L}_{G}), 
\end{aligned} 
\end{equation}
}{
\begin{equation}
\label{eq:QGstep}
\begin{aligned}
Q \leftarrow Q - \eta \nabla_Q (\mathcal{L}_{Q} + \mathcal{L}_{G}), \quad G \leftarrow G - \eta \nabla_G (\mathcal{L}_{Q} + \mathcal{L}_{G}), 
\end{aligned} 
\end{equation}
}
where $\eta > 0$ is the learning rate.

\textbf{F-step}: Fixing $G(\cdot, \cdot)$ and $Q(\cdot, \vecn)$, we optimize $F(\cdot)$ to learn to extract identical RFFs from the raw signals and the corresponding augmented signals with different backgrounds, as presented in Fig.~\ref{fig:diagram}. Similar to (\ref{eq:QGstep}), $F(\cdot)$ and the auxiliary classifier are updated as
\ifthenelse{\boolean{isdouble}}{
\begin{equation}
\begin{aligned}
F &\leftarrow F - \eta \nabla_F (\mathcal{L}_{F}), \\
\W &\leftarrow \W - \eta \nabla_\W (\mathcal{L}_{F}),
\end{aligned} 
\end{equation}
}{
\begin{equation}
\begin{aligned}
F \leftarrow F - \eta \nabla_F (\mathcal{L}_{F}), \quad
\W \leftarrow \W - \eta \nabla_\W (\mathcal{L}_{F}),
\end{aligned} 
\end{equation}
}
respectively.
%

The training of the proposed DR learning framework is performed by implementing these two steps iteratively. The corresponding training algorithm is also described in Algorithm~\ref{alg}. 

As the learning progresses, the RFF extractor $F(\cdot)$ is gradually trained to extract only device-relevant information. Learning the background extractor $Q(\cdot, \cdot)$ relies on $F(\cdot)$, and therefore  also benefits from the improvement of $F(\cdot)$. The improvement of $Q(\cdot, \cdot)$ then leads to clearer signal backgrounds, containing less device-relevant information and providing higher quality disentangled representations. With higher quality signal representations, the synthetic signal generator can create more realistic signals that only swap the background with minimal leakage of device-relevant information. The more realistic the augmented signals, the better $F(\cdot)$ can be generalized to real-world unknown channel statistics. 

\ifthenelse{\boolean{isdouble}}{
\begin{algorithm}[t]
  \caption{Proposed DRL for RFF Extraction}
  \label{alg}
\begin{algorithmic}
  \STATE {\bfseries Input:} Training data set $\mathcal{T}$, Batch size $B$.
  \STATE {\bfseries Output:} $F^*$, $Q^*$, and, $G^*$.
  \STATE {\bfseries Hyperparam:} Learning rate $\eta$, radius $\delta$, coefficients $\lambda$, $\alpha$ and $\beta$.
  \REPEAT
  \STATE \# \textbf{Q/G-step}:
  \STATE Draw batch data $(\vecx^{(i)}, \vecy^{(i)})$ from $\mathcal{T}$;
  \STATE Sample $\vecn^{(i)} \sim \mathcal{N}(0, \mathbf{I})$;
  \STATE Compute $\overline{\mathbf{\vecx}}^{(i)} = Q(\vecx^{(i)},\vecn^{(i)})$, $\vecz^{(i)} = F({\mathbf{\vecx}}^{(i)})$;
  \STATE Compute $\mathcal{L}_Q =\mathcal{L}_\text{v} + \lambda \mathcal{L}_\text{p}$ according to (\ref{eq:lb})-(\ref{eq:pbl_Q});
  \STATE Compute $\hat{\mathbf{\vecx}}^{(i)} = G(\vecz^{(i)}, \overline{\mathbf{\vecx}}^{(i)})$;
  \STATE Compute $\mathcal{L}_G$ according to (\ref{eq:pbl_G});
  
  \STATE Update $Q \leftarrow Q - \eta \nabla_Q (\mathcal{L}_Q + \mathcal{L}_G)$;

  \STATE Update $G \leftarrow G - \eta \nabla_G (\mathcal{L}_Q + \mathcal{L}_G)$;
  \STATE \# \textbf{F-step}:
  \STATE Draw another batch $(\vecx^{(j)}, \vecy^{(j)})$ from $\mathcal{T}$. Sample $\vecn^{(j)} \sim \mathcal{N}(0, \mathbf{I})$;
  \STATE Compute $\overline{\mathbf{\vecx}}^{(j)} = Q(\vecx^{(j)},\vecn^{(j)})$;
  \STATE Swap the background and generate $\hat{\mathbf{\vecx}}^{(i, j)} = G(\vecz^{(i)}, \overline{\mathbf{\vecx}}^{(j)})$ ;
  \STATE Compute $\mathcal{L}_{F}$ according to (\ref{eq:pbl_F});
  \STATE Update $F \leftarrow F - \eta \nabla_F \mathcal{L}_F$;
  \STATE Update $\mathbf{W} \leftarrow \mathbf{W} - \eta \nabla_\mathbf{W} \mathcal{L}_{\text{RFF}}$;
  \UNTIL{convergence}
  \STATE \textbf{return} $F$, $Q$, and $G$.
\end{algorithmic}
\end{algorithm}
}{
\vspace{-0.5cm}
\begin{center}
\resizebox{0.9\linewidth}{!}{
\begin{minipage}[t]{1.2\linewidth}
\begin{algorithm}[H]
  \caption{Proposed DRL for RFF Extraction}
  \label{alg}

\begin{algorithmic}
  \STATE {\bfseries Input:} Training data set $\mathcal{T}$, Batch size $B$.
  \STATE {\bfseries Output:} $F^*$, $Q^*$, and, $G^*$.
  \STATE {\bfseries Hyperparam:} Learning rate $\eta$, radius $\delta$, coefficients $\lambda$, $\alpha$ and $\beta$.
  \REPEAT
  \STATE \# \textbf{Q/G-step}:
  \STATE Draw batch data $(\vecx^{(i)}, \vecy^{(i)})$ from $\mathcal{T}$, sample $\vecn^{(i)} \sim \mathcal{N}(0, \mathbf{I})$;
  \STATE Compute $\overline{\mathbf{\vecx}}^{(i)} = Q(\vecx^{(i)},\vecn^{(i)})$, $\vecz^{(i)} = F({\mathbf{\vecx}}^{(i)})$;
  \STATE Compute $\mathcal{L}_Q =\mathcal{L}_\text{v} + \lambda \mathcal{L}_\text{p}$ according to (\ref{eq:lb})-(\ref{eq:pbl_Q});
  \STATE Compute $\hat{\mathbf{\vecx}}^{(i)} = G(\vecz^{(i)}, \overline{\mathbf{\vecx}}^{(i)})$;
  \STATE Compute $\mathcal{L}_G$ according to (\ref{eq:pbl_G});
  \STATE Update $Q \leftarrow Q - \eta \nabla_Q (\mathcal{L}_Q + \mathcal{L}_G)$, $G \leftarrow G - \eta \nabla_G (\mathcal{L}_Q + \mathcal{L}_G)$;
  \STATE \# \textbf{F-step}:
  \STATE Draw another batch $(\vecx^{(j)}, \vecy^{(j)})$ from $\mathcal{T}$,  sample $\vecn^{(j)} \sim \mathcal{N}(0, \mathbf{I})$;
  \STATE Compute $\overline{\mathbf{\vecx}}^{(j)} = Q(\vecx^{(j)},\vecn^{(j)})$;
  \STATE Swap the background and generate $\hat{\mathbf{\vecx}}^{(i, j)} = G(\vecz^{(i)}, \overline{\mathbf{\vecx}}^{(j)})$;
  \STATE Compute $\mathcal{L}_{F}$ according to (\ref{eq:pbl_F});
  \STATE Update $F \leftarrow F - \eta \nabla_F \mathcal{L}_F$, $\mathbf{W} \leftarrow \mathbf{W} - \eta \nabla_\mathbf{W} \mathcal{L}_{\text{RFF}}$;
  \UNTIL{convergence}
  \STATE \textbf{return} $F$, $Q$, and $G$.
\end{algorithmic}
\end{algorithm}
\end{minipage}
}
\end{center}
}
\ifthenelse{\boolean{isdouble}}{
\begin{table}[!t]  
\caption{The Basic Structure of the RFF extractor $F(\cdot)$}
\centering
\resizebox{\linewidth}{!}{
\begin{tabular}{clc}  

\toprule
\multicolumn{3}{l}{{\bf HyperParams}: Image width $s$, complexity $L$} \\  
\midrule  
\multicolumn{3}{l}{{\bf Input}: Signal $\vecx \in \mathbb{C}^{M}$ $\rightarrow$ Image $\mathbf{I} \in \mathbb{R}^{2\times \frac{M}{S} \times S}$} \\  
\midrule
\multicolumn{3}{l}{\bf Convolution layers}  \\ 
  {\bf Layers} & {\bf Parameters} & {\bf Activation}\\
  $i$   & {\bf Filters:} $2^{i-1}L\times 3\times3$   & BN + $\text{LReLU}_{(0.2)}$   \\ 
      & {\bf Stride:}  $2 - (i\mod 2)$   &   \\ 
      & {\bf Padding:} $1$   &   \\ 
\multicolumn{3}{l}{{\footnotesize Applying convolutional layers until the output is smaller than the filter size.}}\\
\midrule
\multicolumn{3}{l}{{\bf Output}: FC(output of convolutional layers, output dimension)}\\
\bottomrule  

\end{tabular}
}
\label{tb:nn}
\end{table}
}{
\begin{table}[!t]  
\caption{The Basic Structure of the RFF extractor $F(\cdot)$}
\centering
\resizebox{0.5\linewidth}{!}{
\begin{tabular}{clc}  

\toprule
\multicolumn{3}{l}{{\bf HyperParams}: Image width $s$, complexity $L$} \\  
\midrule  
\multicolumn{3}{l}{{\bf Input}: Signal $\vecx \in \mathbb{C}^{M}$ $\rightarrow$ Image $\mathbf{I} \in \mathbb{R}^{2\times \frac{M}{S} \times S}$} \\  
\midrule
\multicolumn{3}{l}{\bf Convolution layers}  \\ 
  {\bf Layers} & {\bf Parameters} & {\bf Activation}\\
  $i$   & {\bf Filters:} $2^{i-1}L\times 3\times3$   & BN + $\text{LReLU}_{(0.2)}$   \\ 
      & {\bf Stride:}  $2 - (i\mod 2)$   &   \\ 
      & {\bf Padding:} $1$   &   \\ 
\multicolumn{3}{l}{{\footnotesize Applying convolutional layers until the output is smaller than the filter size.}}\\
\midrule
\multicolumn{3}{l}{{\bf Output}: FC(output of convolutional layers, output dimension)}\\
\bottomrule  

\end{tabular}
}
\label{tb:nn}
\vspace{-0.5cm}
\end{table}
}

\subsection{Implementation Details}
We propose to adopt CNNs to learn the representations. Unless otherwise specified, the implementation of the proposed DR learning framework uses the following settings:
\begin{itemize}
	\item \textbf{\emph{Preprocessing}}. All input signals to the neural networks are first normalized to $[-1, 1]$ and then converted into images as in our previous work~\cite{xie2021generalizable}. \bflag{Specifically, the 1280-length preamble signal contains eight identical symbols. We convert the signal into 2-channel real-valued images of dimension $(2 \times 16 \times 80)$ such that each row in the image corresponds to one-half of the symbol period. This corresponds to the use of 16 chips in the IEEE 802.15.4 standard, and thus we have a total of 80 sample points for a 1280-length preamble. This way, the pixels in a row are from the same symbol, whereas pixels from disjoint rows belong to different symbols.}
	 \bflag{The impact of the hardware characteristics and wireless channels in the received signals are therefore reflected as texture and edge differences in the images, facilitating the subsequent learning by the convolutional layers.}
	\item \textbf{\emph{The RFF extractor.}} The RFF extractor, $F(\cdot)$, is implemented using the basic convolutional neural network~(BCNN) adopted in \cite{xie2021generalizable}, as shown in Table~\ref{tb:nn}. We employ a small filter with  few parameters in the convolutional layers to achieve a large effective receptive field. Batch normalization and LeakyReLU(0.2) are adopted for training stability and network non-linearity, respectively. We continue applying convolutional layers until the output feature maps are smaller than the filter size, i.e., $(3\times 3)$, and the final output representations are computed by a fully connected layer. The hyper-parameter $L$ controls the network complexity.
	\item \textbf{\emph{The background extractor \& the signal generator}}. For domain \bflag{and background information} preservation, the background signal extractor $Q(\cdot, \vecn)$ and the signal generator $G(\cdot, \cdot)$ in this work are both implemented using a U-net~\cite{ronneberger2015u}. The U-net is a specific type of CNN with symmetrical shortcuts designed for image-domain-preserving processing and is widely used in \bflag{high-fidelity} medical image processing and image segmentation. The detailed structure of $Q(\cdot,\vecn)$ and $G(\cdot, \cdot)$ is discussed in Appendix A.
	\item \textbf{\emph{Optimizer}}. All of the neural networks are trained using Adam~~\cite{kingma2014adam} with learning rate $\eta=0.001$ and parameters $\beta_1=0.9, \text{ and }\beta_2=0.999$. 
	\item \textbf{\emph{Hyper-parameters}}. \bflag{The proposed approach works well in our experimental test sets for a wide range of the parameters: $\lambda \in [0.3, 0.6]$,  $\alpha \in [5, 50]$, $\beta \in [5, 50]$~(see Section IV.D).} In the results depicted in the next section, we set the hyper-parameters as $\lambda = 0.5$, $\alpha = 10$, and $\beta=10$. We also set the hyper-parameter in the information constraint in (\ref{eq:bg}) to be $0$, i.e., $\epsilon=0$. As in the previous work~\cite{ranjan2017l2}, we set the radius of the HP to be $\delta=10$.
\end{itemize}

The implementation details \bflag{and datasets} are also available online in our Github at \cite{xie2021DR}. All the source codes are implemented in PyTorch using our own research toolbox {\bf MarvelToolbox}~\cite{xie2019Marval}.

\ifthenelse{\boolean{isdouble}}{
\begin{figure}[t]
	\centering
	\includegraphics[width=0.99\linewidth]{\rootpath/figures_pdf/layout.pdf}
	\caption{The layout of device positions in the testbed. }
	\label{fig:dataset}
\end{figure}
}{
\begin{figure}[t]
	\centering
	\includegraphics[width=0.7\linewidth]{\rootpath/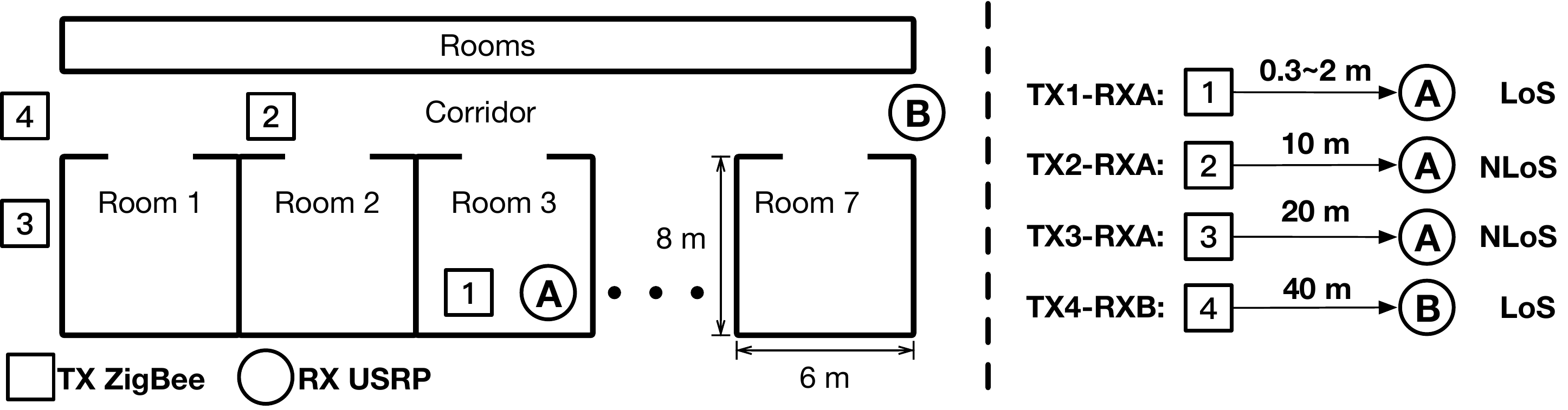}
	\caption{\bflag{The layout of device positions in the testbed.}}
	\label{fig:dataset}
	\vspace{-0.5cm}
	\end{figure}
}

\section{Experimental Evaluation}
In this section, \bflag{we evaluate the effectiveness of the proposed DRL framework using data collected from a real-world testbed.} \bflag{We compare the performance of the proposed DR-RFF extractor with that of a typical closed-set RFF classifier, ML RFF extractor, and the ML RFF extractor trained with different DA methods.} The experiments consist of four parts: 1) Performance comparisons for different open test sets which contain both the unknown devices and the unknown multi-path channels; 2) Performance comparison for different signal-to-noise ratios~(SNRs); \bflag{3) Hyper-parameters tuning;} 4) Learning curve comparisons for overfitting evaluation.

\subsection{Experimental Setup} 
\paragraph{Dataset} We exploit the signals transmitted from 59 TI CC2530 ZigBee devices and collected via a USRP N210 receiver in different positions. All ZigBee devices operate at 2.4 GHz with a maximum transmit power of 19 dBm. The sampling rate of the receiver is 10 Msample/s and thus each preamble signal $\vecx$ contains $M=1280$ sample points.

To evaluate the effectiveness of the proposed DR learning framework under the unknown channel statistics, we collect the required datasets form the different positions shown in the left-hand side of Fig.~\ref{fig:dataset}. We denote the signals collected from the ZigBee devices transmitting at position 1 and received at position A as {\bf TX1-RXA}. \bflag{Analogously, we depict four collecting positions in the right-hand side of Fig. \ref{fig:dataset}.} \bflag{Note that the signals collected from the 54 ZigBee devices in {\bf TX1-RXA} were used for evaluating closed-set classification performance in~\cite{peng2018design, yu2019robust}, and for evaluating the performance of open-set scenarios in our previous work~\cite{xie2021generalizable}.}
Table \ref{tb:dataset} provides further details about the data sets used in this paper. 
The training and validation sets contain signals from 45 ZigBee devices under TX1-RXA collected in 2016. The test sets can be divided into types according to whether they have the same propagation environment with the training set: 
\begin{itemize}
\item {\bf T1-T3} are test sets collected in the same propagation environment as the training set, and the algorithm performance is evaluated based on whether 1) the test sets contain known devices and 2) they experience device aging. The devices considered in T2 and T3 experienced device aging since they operated continuously for over 18 months.
\item {\bf M1-M3} are collected from five unknown devices and three types of unknown wireless channels in order of classification difficulty from easy to hard. The easiest one, i.e., M1, contains only a single unknown multi-path fading channel, while M3 has three types of unknown channels and is the most challenging case considered for open-set classification.
\end{itemize}
{Since the test sets are collected with different positions and running times, the main factors that affect the identification performance are the unknown multi-path fading channels and device aging.}
  
\begin{table*}[t]  

\caption{Dataset for Evaluation}

\centering
\resizebox{0.8\linewidth}{!}{	
\begin{tabular}{c|c|c|c|ccc}  
\toprule   
\multirow{2}*{Datasets} & \multirow{2}*{Device IDs} & \multicolumn{2}{c|}{Collection Environment} & \multicolumn{3}{c}{Properties} \\  
\cline{3-7}
~ & ~& Positions & Dates & Unknown Device & Device Aging & Multi-path\\
\midrule
\hline
Training set & \multirow{2}*{1-45} & \multirow{2}*{TX1-RXA} & \multirow{2}*{Jun. 2016} & - & - & $\times$ \\
\cline{1-1}
\cline{5-7}
Validation set & ~ & ~ & ~ & $\times$ & $\times$ & $\times$ \\
\midrule
\multirow{1}*{Test set: T1} & \multirow{1}*{46-54} & \multirow{3}*{TX1-RXA} & Jun. 2016 & \multirow{1}*{\checkmark} & \multirow{1}*{$\times$} & \multirow{1}*{$\times$} \\
\cline{1-2}
\cline{4-7}
\multirow{1}*{Test set: T2} & \multirow{1}*{1-45} & ~ & Jan. 2018, & \multirow{1}*{$\times$} & \multirow{1}*{\checkmark} & \multirow{1}*{$\times$} \\
\cline{1-2}
\cline{5-7}
Test set: T3 & \multirow{1}*{46-54} & ~ & Feb. 2018 &  \checkmark  & \multirow{1}*{\checkmark} & $\times$ \\
\midrule
Test set: M1 & \multirow{4}*{55-59} &  TX2-RXA & \multirow{4}*{Apr. 2018} & \checkmark & $\times$ & \checkmark \\
\cline{1-1}
\cline{3-3}
\cline{5-7}
{Test set: M2} & ~ & TX2-RXA, TX3-RXA & ~ & {\checkmark} & {$\times$} & {\checkmark} \\
\cline{1-1}
\cline{3-3}
\cline{5-7}
\multirow{2}*{Test set: M3} & ~ & TX2-RXA, TX3-RXA, & ~ & \multirow{2}*{\checkmark} & \multirow{2}*{$\times$} & \multirow{2}*{\checkmark} \\
~ & ~ &TX4-RXB & ~ \\
\bottomrule  
\end{tabular}
}
\label{tb:dataset}
\vspace{-0.5cm}
\end{table*}

\paragraph{Metrics}
As commonly adopted in the open-set recognition tasks~\cite{danev2012physical,xie2021generalizable, deng2019arcface}, we use the receiver operating characteristic~(ROC) curve, the area under the ROC curve (AUC), and the equal error rate (EER) operating point to evaluate the performance of the RFF extractors. The ROC curve depicts the trade-off between true-positive rate~(TPR) and false-positive rate~(FPR). To obtain the ROC curve, we compute the TPR and the FPR by traversing the verification thresholds $T$ in (\ref{eq:verification}). Given a certain $T$, TPR refers to the probability that signal pairs from being from the same device are correctly verified as the same devices by the verification system. FPR refers to the percentage of the signal pairs from the same device that yielded false alarms by the verification system. The EER refers to the point where FNR~(i.e., 1-TPR) and FPR are equal. 
A larger AUC and a lower EER indicate a more discriminative RFF that simultaneously achieves fewer false negatives and fewer false positives.

\paragraph{Baselines and the Proposed DR-RFFs} 

\bflag{We consider five categories, a total of eight baseline approaches as summarized in Table \ref{tb:baseline}. They are also listed here as follows:
\begin{itemize}
	\item A typical closed-set RFF classifier, i.e., \textbf{Yu et al.}~\cite{yu2019robust};
	\item A discriminative RFF extractor without data augmentation, i.e., \textbf{ML-RFF}~\cite{xie2021generalizable};
	\item Handcrafted data augmentation, i.e.,  \textbf{AWGN}~\cite{yu2019multi} and \textbf{FIR}~\cite{soltani2020more};
	\item Learning-based data augmentation, i.e., \textbf{PGD} adversarial training~\cite{madry2017towards};
	\item The proposed method, i.e., \textbf{DR-RFF}, and its two types of variants for ablation study.
\end{itemize} 
Except for \textbf{Yu et al.}~\cite{yu2019robust}, all baseline approaches are the discriminative RFF extractor proposed in \cite{xie2021generalizable}, but with different data augmentation methods. }
Unless otherwise specified, all the baseline approaches use the same network structure~(see Table~\ref{tb:nn}) with the same complexity setting of $L=18$. All algorithms are trained by Adam with the same setting as the proposed DR-RFF. \bflag{To better assess performance, we trained each method ten times and calculate the average performance as well as the corresponding standard deviations.}

\begin{table*}[t]  
\caption{\bflag{Baselines RFF Extractors and The Proposed DR-RFF}}

\centering
\resizebox{\linewidth}{!}{
\begin{threeparttable}[b]

\begin{tabular}{l|c|c|c|c|c}  

\toprule 
\multirow{2}*{Baselines}&\multirow{2}*{Training methods}&\multirow{2}*{Data augmentation} &\multicolumn{3}{c}{Conditional distribution in (4):  $p_\mathbf{W}(\mathbf{y}|F(\mathbf{x}))$}\\
\cline{4-6}
 ~  & ~& ~&RFF extractor & Auxiliary classifier & \multirow{1}*{\# Parameters}\\  
\midrule
\hline
\multirow{1}*{\bflag{Yu et al.~\cite{yu2019robust}}} & \multirow{5}*{MLE} & AWGN~($\text{SNR: } 5 \sim 30 $ dB) & \multirow{7}*{BCNN~\cite{xie2021generalizable}} & Softmax & \multirow{8}*{Approx. $7$ M} \\
\cline{1-1}
\cline{3-3}
\cline{5-5}
\multirow{1}*{ML-RFF~\cite{xie2021generalizable}} & ~ & \multirow{1}*{N/A} & ~ & \multirow{5}*{Softmax with HP~\cite{xie2021generalizable}} & ~ \\
\cline{1-1}
\cline{3-3}
\multirow{1}*{AWGN~\cite{yu2019robust}} & ~ & AWGN~($\text{SNR: } 5 \sim 30 $ dB) &~& ~ & ~ \\
\cline{1-1}
\cline{3-3}
\multirow{1}*{\bflag{FIR~\cite{soltani2020more}}}& ~ & Gaussian FIR filtering (9 taps) & ~ & ~ & ~ \\
\cline{1-1}
\cline{3-3}
\multirow{1}*{\bflag{PGD~\cite{madry2017towards}}} & ~ &  PGD attack~($l_\infty$-norm bounded by 0.1) & ($L=18$) & ($\delta = 10$) & ~  \\ 
\cline{1-3}
\multirow{1}*{DR-RFF\tnote{\dag}} & \multicolumn{2}{l|}{\multirow{1}{*}{The proposed framework}} & ~ & ~ & ~ \\
\cline{1-3}
\multirow{1}*{DR-RFF\tnote{\dag} w/o BS} &  \multicolumn{2}{l|}{The proposed framework without background shuffling} &~ & ~ & ~ \\
\cline{1-3}
\cline{5-5}
\multirow{1}*{\bflag{DR-RFF\tnote{\dag} w/o HP}} &  \multicolumn{2}{l|}{The proposed framework without HP} &~ & Softmax & ~ \\

\bottomrule  
\end{tabular}

\begin{tablenotes}
     \item[\dag]Proposed method in this paper.
\end{tablenotes}

\end{threeparttable}
}
\label{tb:baseline}
\vspace{-0.5cm}
\end{table*}

\begin{figure*}[t]
	\centering
	\subfigure[{\scriptsize T1: Unknown devices/no device aging.}]{
        \centering
        \begin{minipage}[t]{0.32\linewidth}
        \centering
        \includegraphics[width=\linewidth]{\rootpath/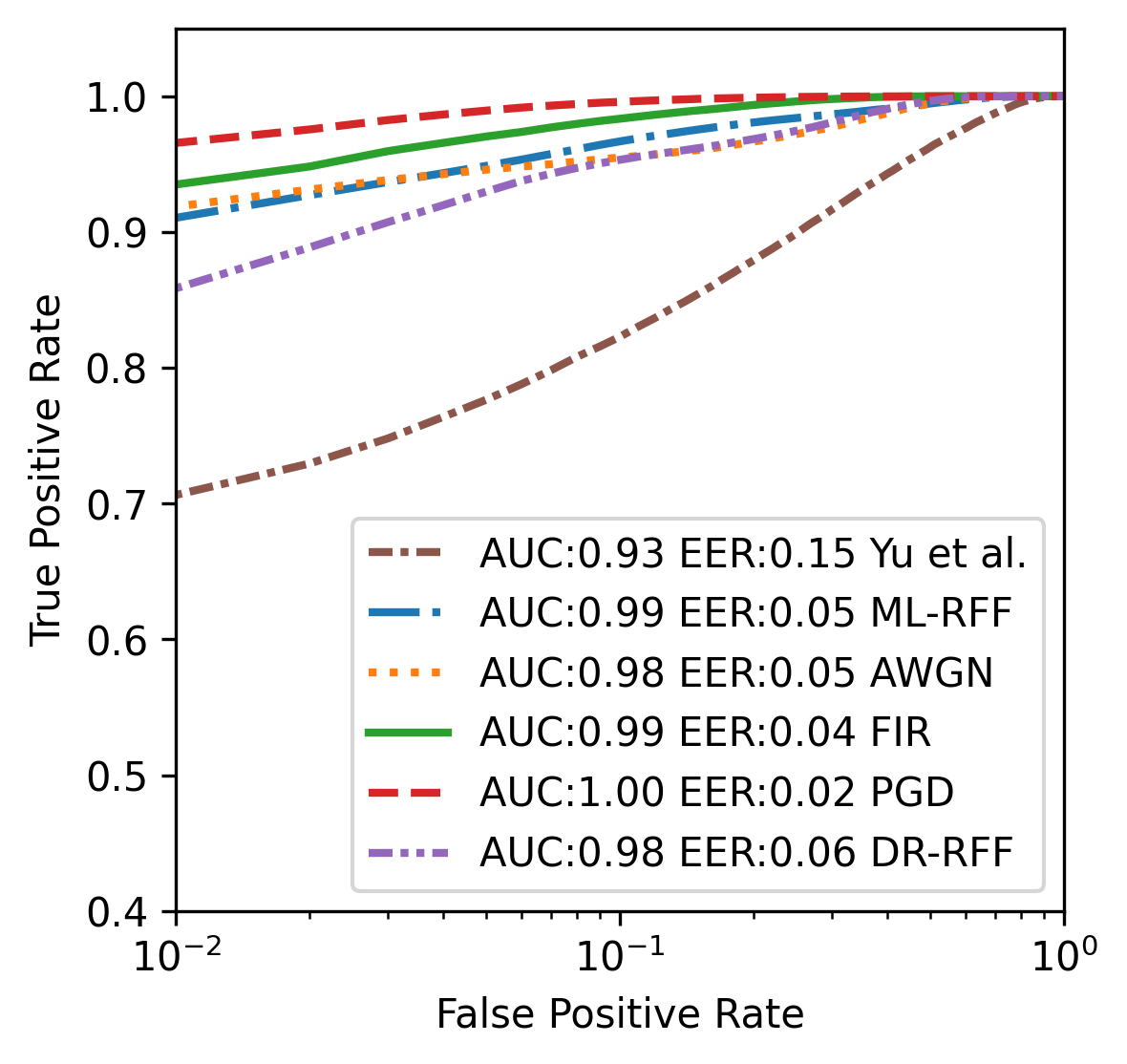}
        \end{minipage}
    }%
	\subfigure[{\scriptsize T2: Known devices/device aging.}]{
        \centering
        \begin{minipage}[t]{0.315\linewidth}
        \centering
        \includegraphics[width=\linewidth]{\rootpath/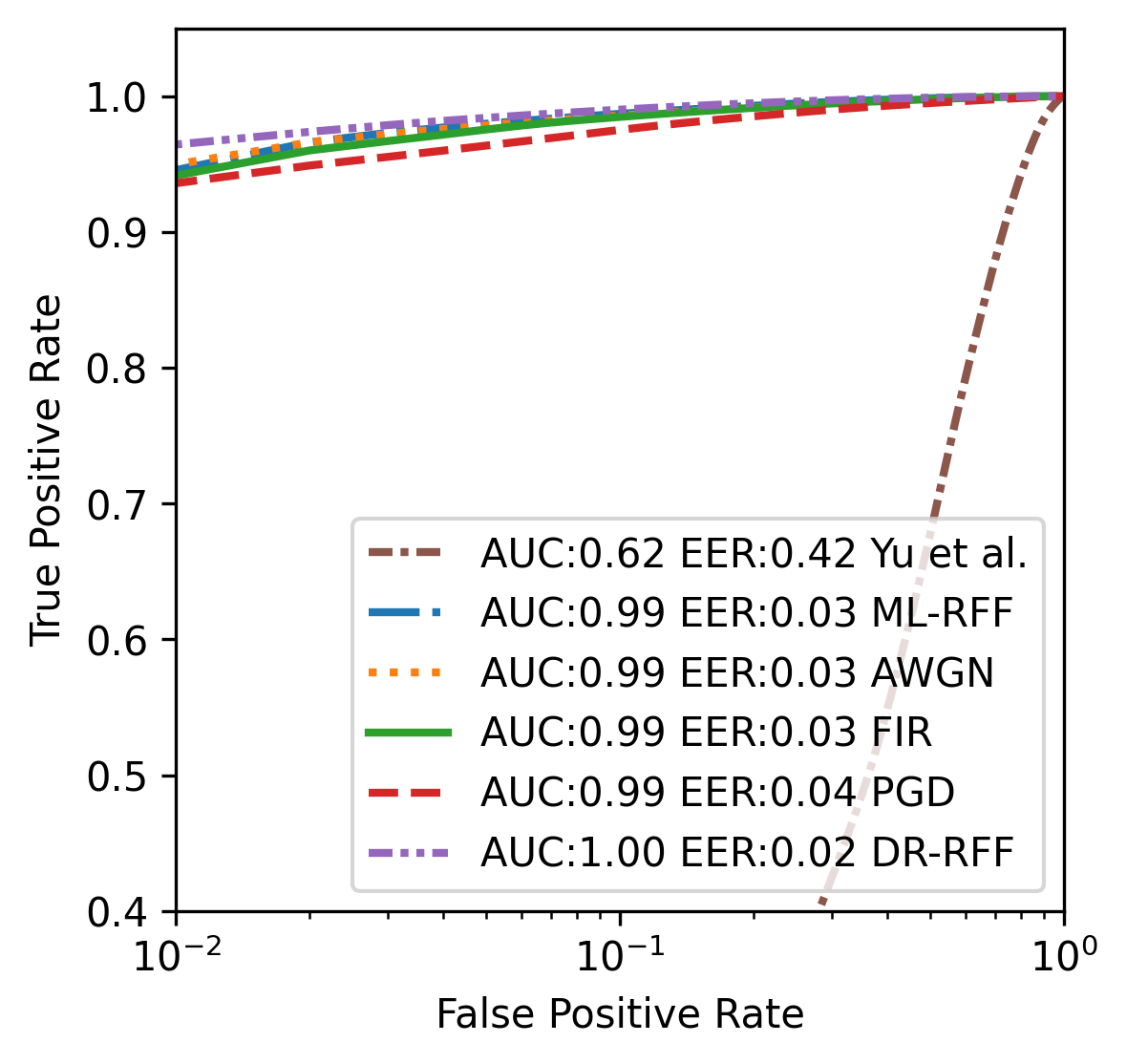}
        \end{minipage}
    }%
    \subfigure[{\scriptsize T3: Unknown devices/device aging.}]{
        \centering
        \begin{minipage}[t]{0.32\linewidth}
        \centering
        \includegraphics[width=\linewidth]{\rootpath/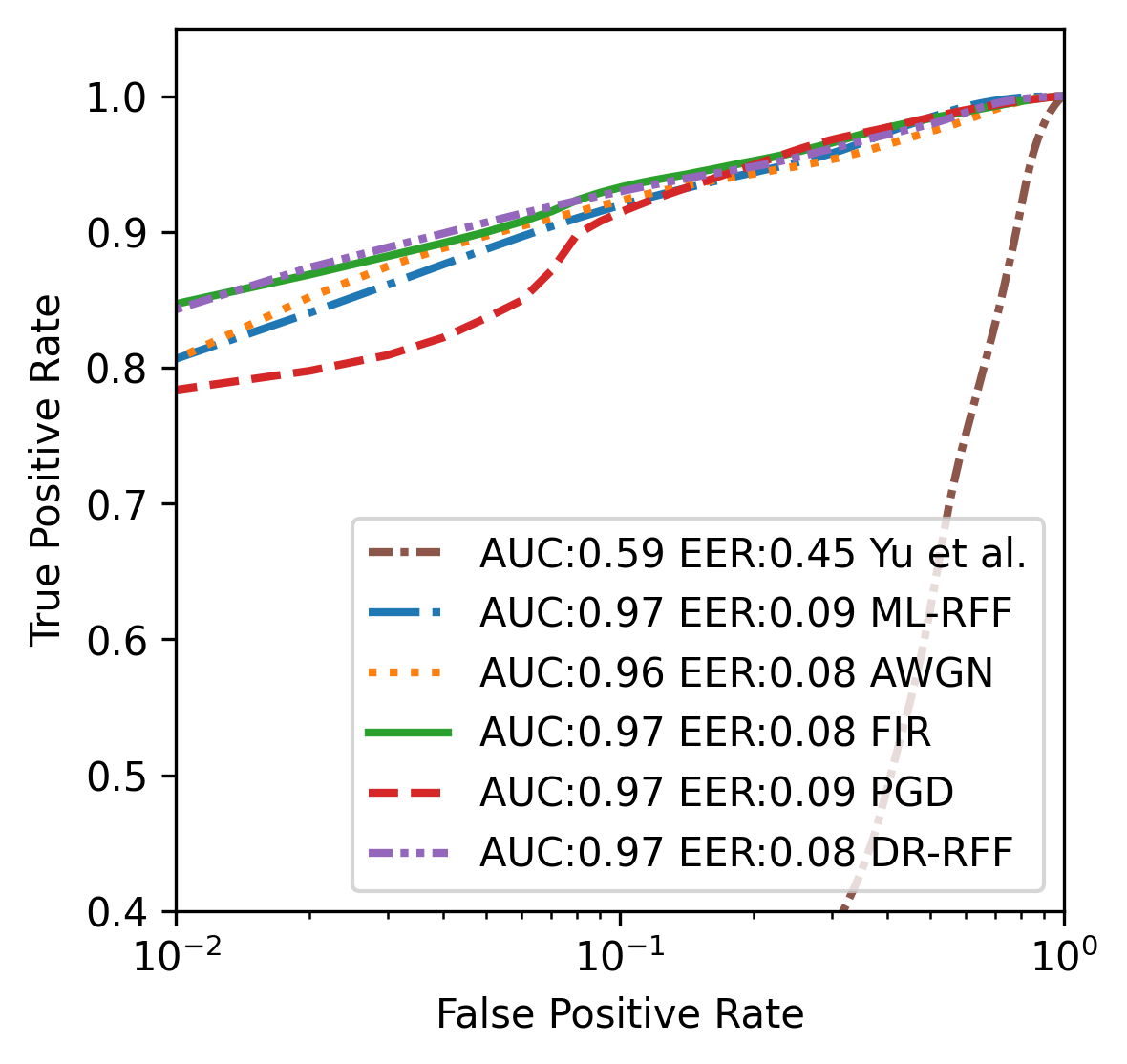}
        \end{minipage}
    }\\
	\subfigure[{\scriptsize M1:Unknown devices/1 unknown channel.}]{
        \centering
        \begin{minipage}[t]{0.32\linewidth}
        \centering
        \includegraphics[width=\linewidth]{\rootpath/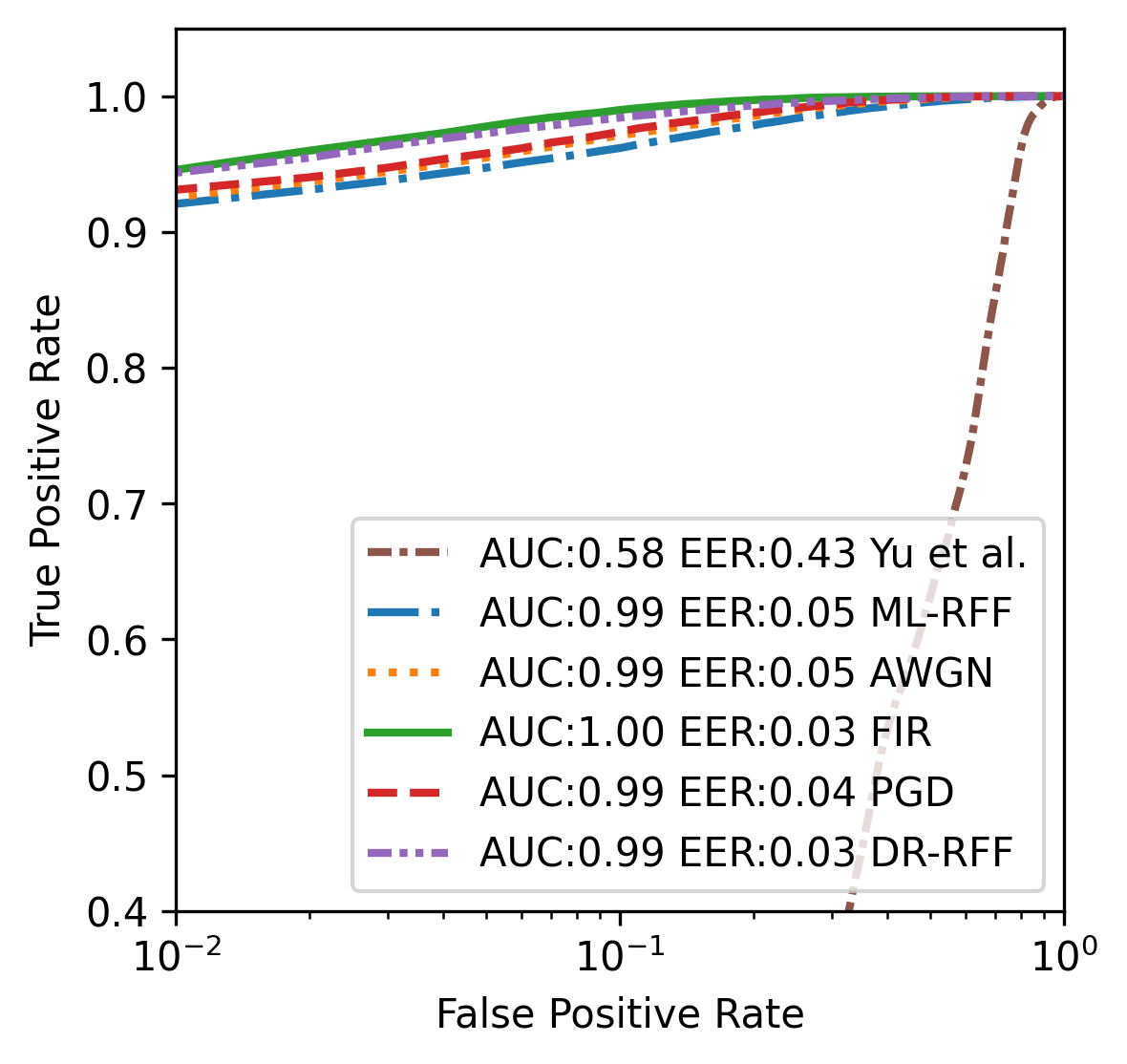}
        \end{minipage}
    }%
    \subfigure[{\scriptsize M2:Unknown devices/2 unknown channels.}]{
        \centering
        \begin{minipage}[t]{0.32\linewidth}
        \centering
        \includegraphics[width=\linewidth]{\rootpath/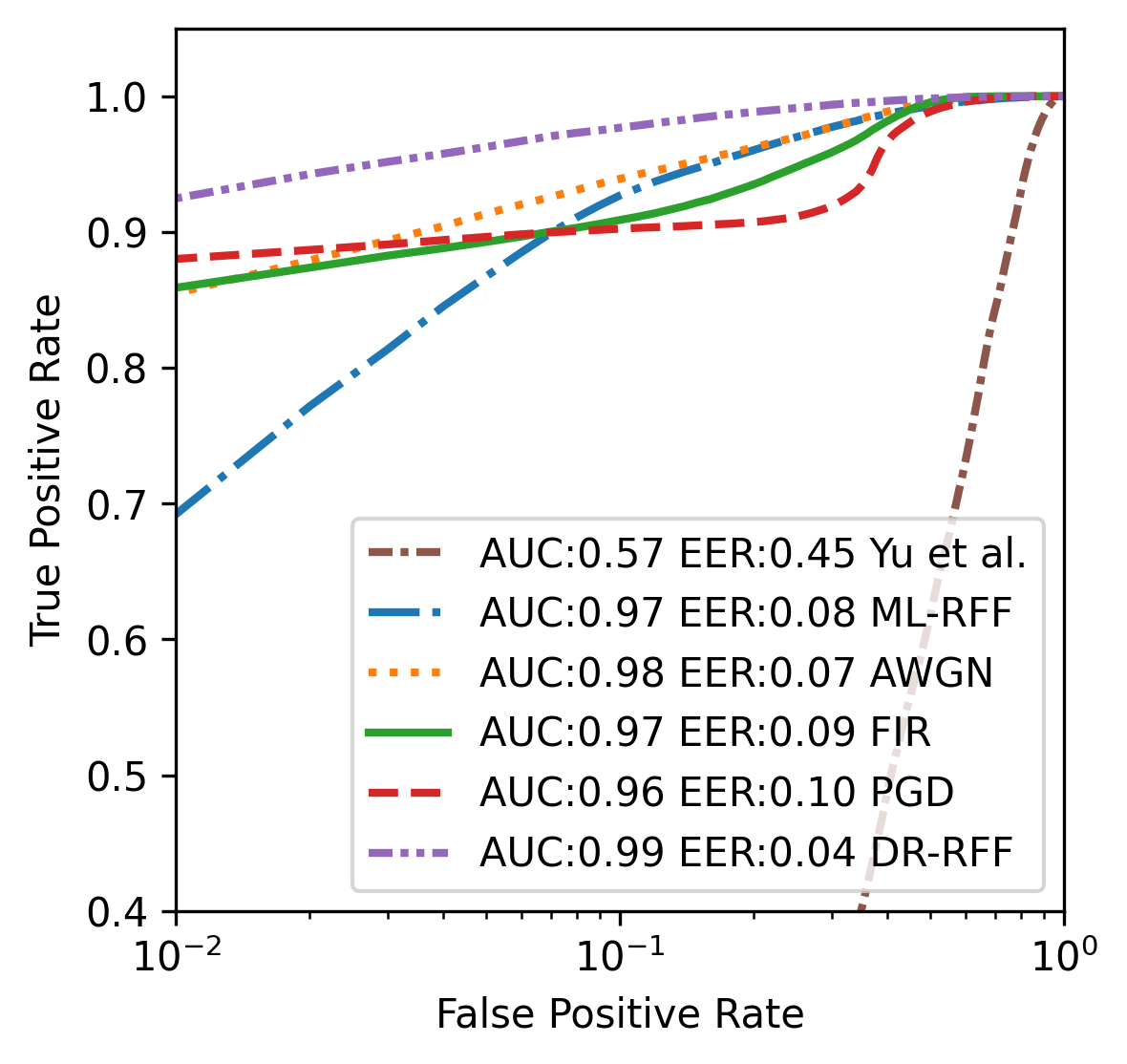}
        \end{minipage}
    }%
    \subfigure[{\scriptsize M3:Unknown devices/all unknown channels.}]{
        \centering
        \begin{minipage}[t]{0.32\linewidth}
        \centering
        \includegraphics[width=\linewidth]{\rootpath/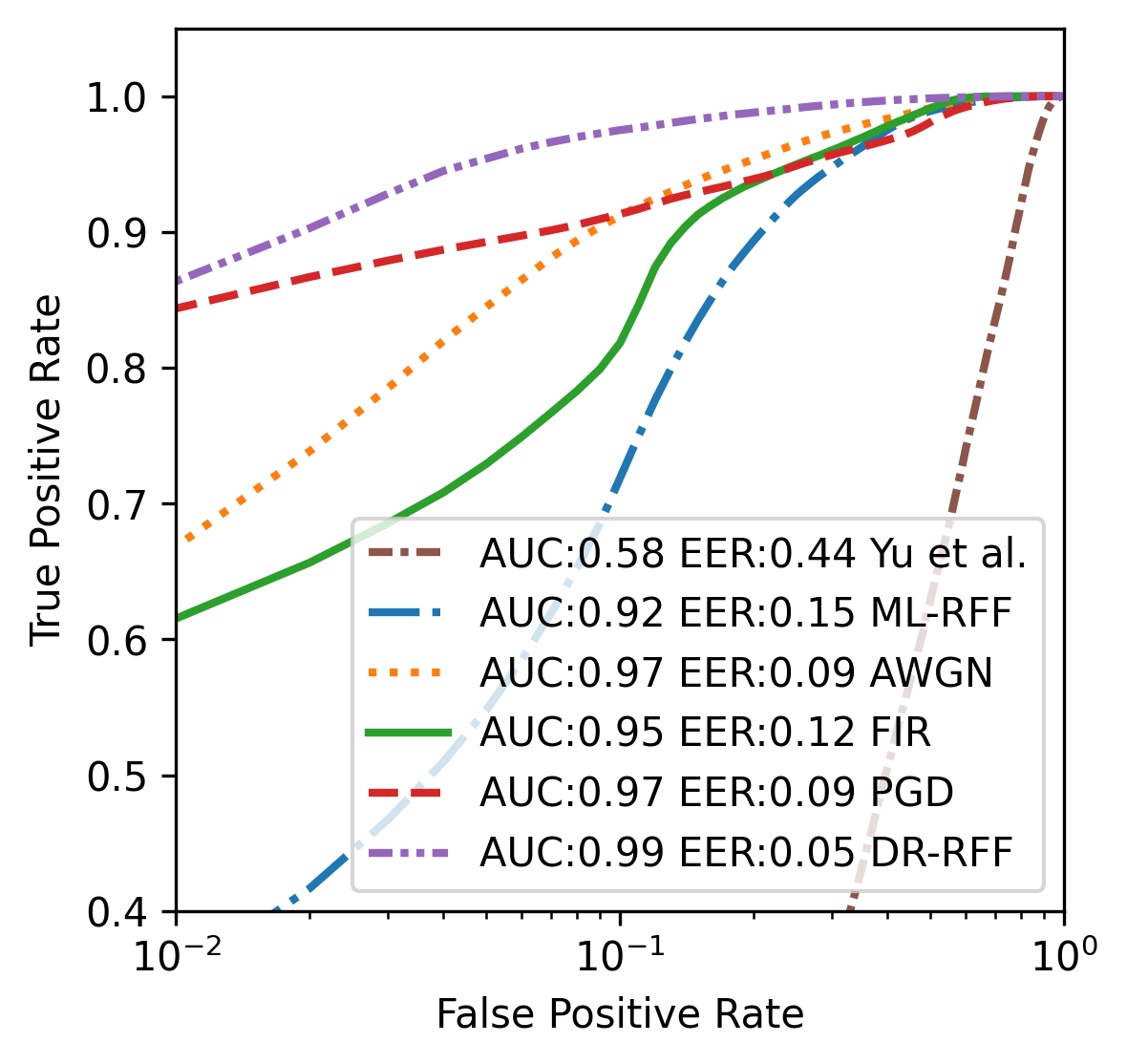}
        \end{minipage}
    }\\
	\caption{\bflag{Average ROC curves of different methods under different open set settings (SNR $\approx$ 30 dB). The test sets, T1-T3, are collected at the same position as the training set, while M1-M3 are collected from unknown devices at unknown positions.}}
	\label{fig:exp0}
	\vspace{-0.5cm}
\end{figure*}

\begin{table*}[htbp]  

\caption{\bflag{ROC Comparison of Different Methods}}

\centering
\resizebox{\linewidth}{!}{
\begin{threeparttable}[b]

\begin{tabular}{llcccccccccccc}  
\toprule   
\multirow{2}*{Baselines} &\multicolumn{2}{c}{T2}&\multicolumn{2}{c}{T3} &\multicolumn{2}{c}{M1}&\multicolumn{2}{c}{M2}&\multicolumn{2}{c}{M3}\\
		~  & AUC(\%) & EER(\%) & AUC(\%) & EER(\%) & AUC(\%) & EER(\%)& AUC(\%) & EER(\%) & AUC(\%) & EER(\%) \\
 
\midrule
\bflag{Yu et al.~\cite{yu2019robust}} & 62.07$_{\pm0.69}$ & 42.14$_{\pm0.71}$& 59.45$_{\pm1.27}$ & 44.63$_{\pm2.46}$& 58.28$_{\pm1.34}$ & 43.38$_{\pm1.91}$& 57.13$_{\pm1.08}$ & 44.65$_{\pm1.05}$& 58.07$_{\pm1.08}$ & 44.12$_{\pm0.81}$\\
\midrule
ML-RFF~\cite{xie2021generalizable} & 99.46$_{\pm0.23}$ & 2.75$_{\pm0.51}$& 96.82$_{\pm0.64}$ & 8.56$_{\pm1.12}$& 98.77$_{\pm0.40}$ & 5.19$_{\pm0.77}$& 97.13$_{\pm0.56}$ & 8.38$_{\pm0.77}$& 92.14$_{\pm1.92}$ & 15.39$_{\pm1.78}$ \\
AWGN~\cite{yu2019robust}   & 99.38$_{\pm0.07}$ & 2.85$_{\pm0.19}$& 96.44$_{\pm0.35}$ & 8.33$_{\pm0.59}$& 99.09$_{\pm0.14}$ & 4.65$_{\pm0.41}$& 97.95$_{\pm0.36}$ & 7.24$_{\pm0.81}$& 96.69$_{\pm0.35}$ & 9.36$_{\pm0.77}$\\
\bflag{FIR~\cite{soltani2020more} }        & 99.32$_{\pm0.06}$ & 3.19$_{\pm0.16}$& {97.16}$_{\pm0.39}$ & {7.71}$_{\pm0.43}$& \cellcolor{lightgray}\bf{99.63}$_{\pm0.10}$ & \cellcolor{lightgray}\bf{3.17}$_{\pm0.50}$& 97.14$_{\pm0.33}$ & 9.30$_{\pm0.65}$& 95.18$_{\pm0.24}$ & 12.23$_{\pm0.67}$\\
\bflag{PGD~\cite{madry2017towards}}   & 99.03$_{\pm0.07}$ & 3.99$_{\pm0.38}$& 96.61$_{\pm0.30}$ & 9.12$_{\pm0.50}$& 99.21$_{\pm0.27}$ & 4.44$_{\pm1.02}$& 96.18$_{\pm0.09}$ & 9.78$_{\pm0.21}$& 96.77$_{\pm0.09}$ & 8.98$_{\pm0.57}$ \\
\midrule
DR-RFF\tnote{\dag}  & \cellcolor{lightgray}\bf{99.62}$_{\pm0.02}$ & \cellcolor{lightgray}\bf{2.39}$_{\pm0.06}$& 96.99$_{\pm0.33}$ & 7.77$_{\pm0.31}$& 99.47$_{\pm0.20}$ & 3.44$_{\pm0.67}$& \cellcolor{lightgray}\bf{99.21}$_{\pm0.24}$ & \cellcolor{lightgray}\bf{4.12}$_{\pm0.63}$& \cellcolor{lightgray}\bf{99.00}$_{\pm0.17}$ & \cellcolor{lightgray}\bf{4.79}$_{\pm0.55}$\\
DR-RFF\tnote{\dag} w/o HP & 98.97$_{\pm0.14}$ & 4.22$_{\pm0.65}$& \cellcolor{lightgray}\bf{97.50}$_{\pm0.28}$ & \cellcolor{lightgray}\bf{7.66}$_{\pm0.39}$& 98.44$_{\pm1.26}$ & 5.74$_{\pm2.28}$& 96.63$_{\pm0.81}$ & 9.17$_{\pm0.68}$& 96.72$_{\pm0.83}$ & 9.34$_{\pm1.26}$\\
\bflag{DR-RFF\tnote{\dag} w/o BS} & 97.55$_{\pm0.93}$ & 6.94$_{\pm1.52}$& 94.90$_{\pm1.20}$ & 11.61$_{\pm1.76}$& 95.58$_{\pm2.09}$ & 10.43$_{\pm3.92}$& 94.10$_{\pm2.52}$ & 12.79$_{\pm3.90}$& 93.58$_{\pm3.36}$ & 13.39$_{\pm4.81}$\\
\bottomrule  
\end{tabular}

\begin{tablenotes}
     \item[\dag] Proposed in this paper.
\end{tablenotes}
\end{threeparttable}
}
\label{tb:resluts}
\vspace{-0.5cm}
\end{table*}

\subsection{Performance Under Unknown Devices \& Channel Statistic}
In order to investigate the effectiveness of the proposed framework, we plot the ROC curves in Fig. \ref{fig:exp0} comparing the performance of the RFFs trained under our proposed framework against the baseline algorithms under different open-set settings. We also compare the AUC and the EER in Table \ref{tb:resluts}.   

\paragraph{Power of Disentangled Representation Learning} Overall, \bflag{the RFF extractors trained by the proposed DR learning framework~(\textbf{DR-RFF}) achieve satisfactory performance for consistent propagation conditions and outperform the conventional methods~(e.g., \textbf{Yu et al.}, \textbf{ML-RFF}, \textbf{AWGN}, \textbf{FIR}) and the adversarial training method~(\textbf{PGD}) for the test sets collected under unknown environments. Note that the closed-set RFF classifier, i.e.,\textbf{Yu et al.}, only performs slightly better than a random guess. These results verify that the RFFs extracted by the proposed approach exhibit stronger generalizability to varying wireless propagation scenarios than the others. Even under the most challenging test set, i.e., M3 in Fig.~\ref{fig:exp0}(f), which contains three types of unknown channel statistics, the \textbf{DR-RFF} trained by the proposed framework can still preserve an AUC over $99\%$. Although the conventional methods perform well with known positions or under a single type of unknown channel, their performance degrades notably when propagation conditions vary, e.g., $3.17\%$ average EER of \textbf{FIR} for M1 degrades to $12.23\%$ for M3. These performance degradations result from the overfitting of the channel statistics in the training set or mismatched prior distributions between the data augmentation and real-world scenarios. With the data-driven DR learning, these challenges can be addressed at least to some extent.
These results demonstrate the superiority of the proposed DR learning framework for robust RFF extraction, especially for changing or unknown wireless channels. }

\paragraph{Ablation experiment} \bflag{To verify the necessity of the HP operation (\ref{eq:normed}) and background shuffling~(BS) in the proposed DR learning framework, we compare the proposed method~(\textbf{DR-RFF}) with its ablation of HP~(\textbf{DR-RFF w/o HP}) and  BS~(\textbf{DR-RFF w/o BS}), respectively. We find that \textbf{DR-RFF} outperforms its ablations, as evidenced by both the AUC and EER values in Table \ref{tb:resluts}. For HP ablation, we suggest that the performance degradations are caused by incomplete disentanglement. HP can exclude device-irrelevant information from RFFs and force the background extractor to extract device-irrelevant information to reconstruct the input signals. We also find that BS is crucial for stable training in the proposed framework. Actually, BS makes the proposed framework act like adversarial training by imposing a strong regularization on the current RFF extractor. Without BS, \textbf{DR-RFF} will degenerate to a baseline of an RFF extractor trained using a typical generative model-based DA method and therefore leads to unsatisfactory performance. }

\begin{figure*}[t]
	\centering
	\subfigure[{\scriptsize M1:Unknown devices/1 unknown channel.}]{
        \centering
        \begin{minipage}[t]{0.32\linewidth}
        \centering
        \includegraphics[width=\linewidth]{\rootpath/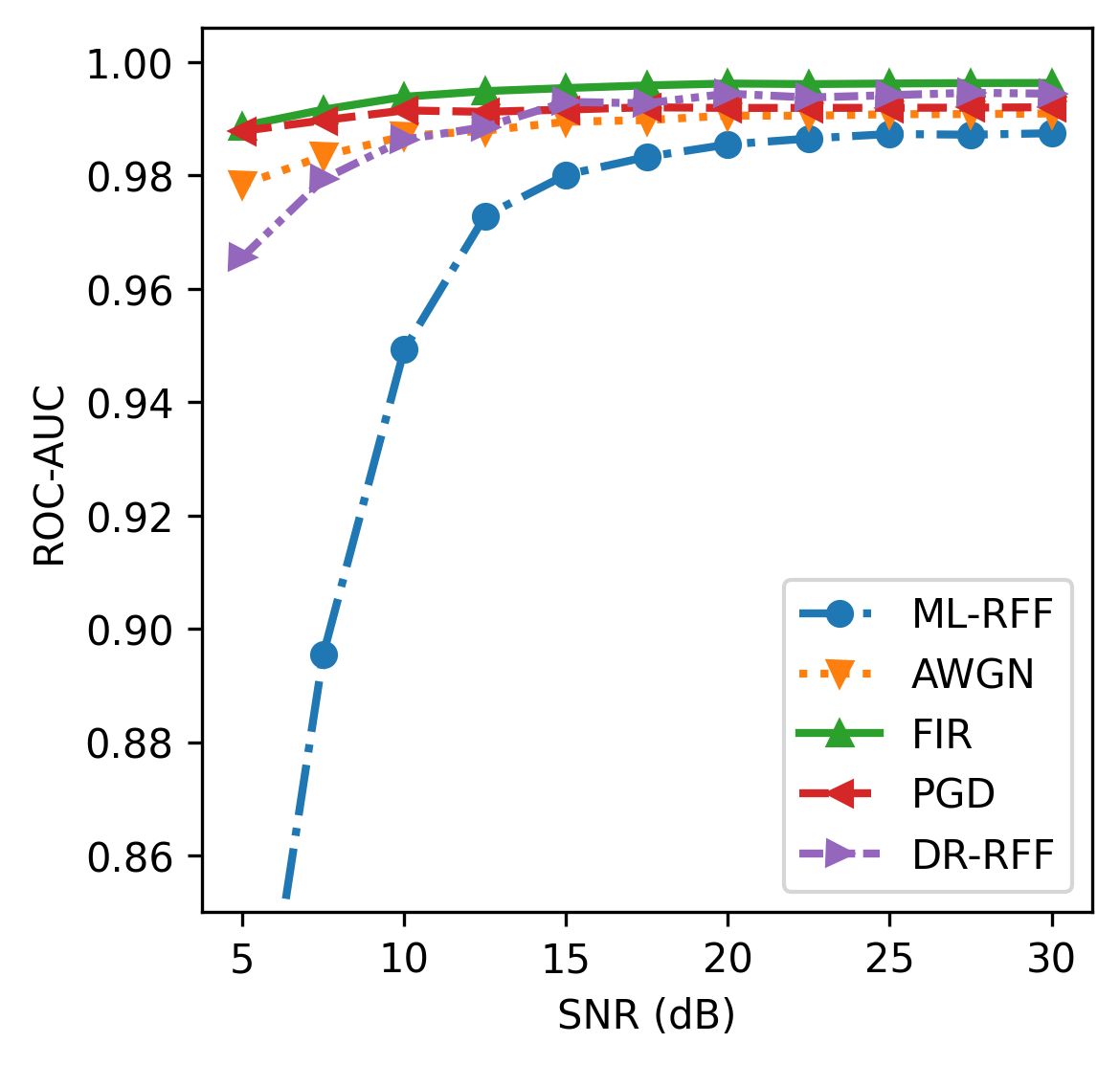}
        \end{minipage}
    }%
    \subfigure[{\scriptsize M2:Unknown devices/2 unknown channels.}]{
        \centering
        \begin{minipage}[t]{0.32\linewidth}
        \centering
        \includegraphics[width=\linewidth]{\rootpath/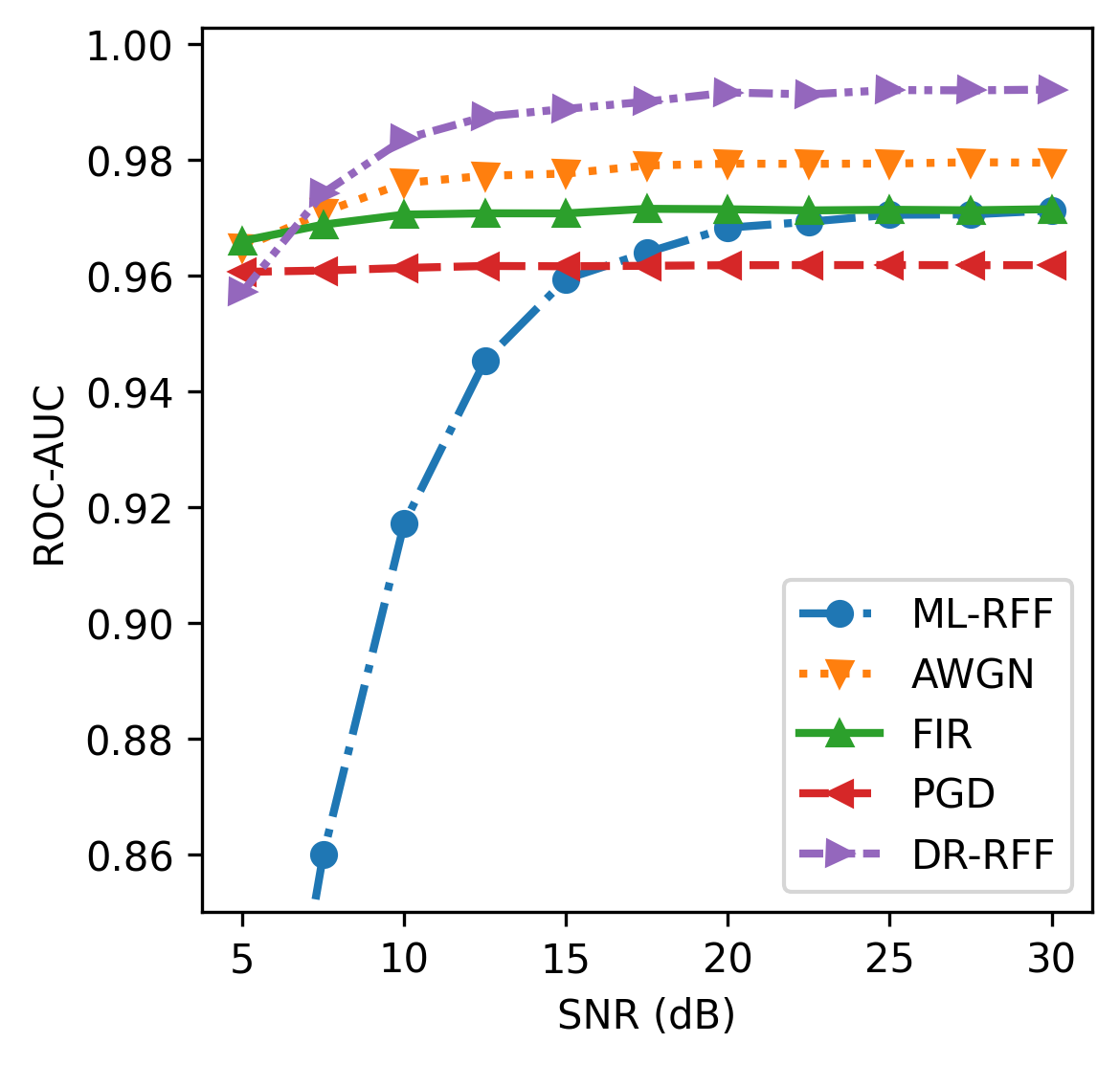}
        \end{minipage}
    }%
    \subfigure[{\scriptsize M3:Unknown devices/all unknown channels.}]{
        \centering
        \begin{minipage}[t]{0.32\linewidth}
        \centering
        \includegraphics[width=\linewidth]{\rootpath/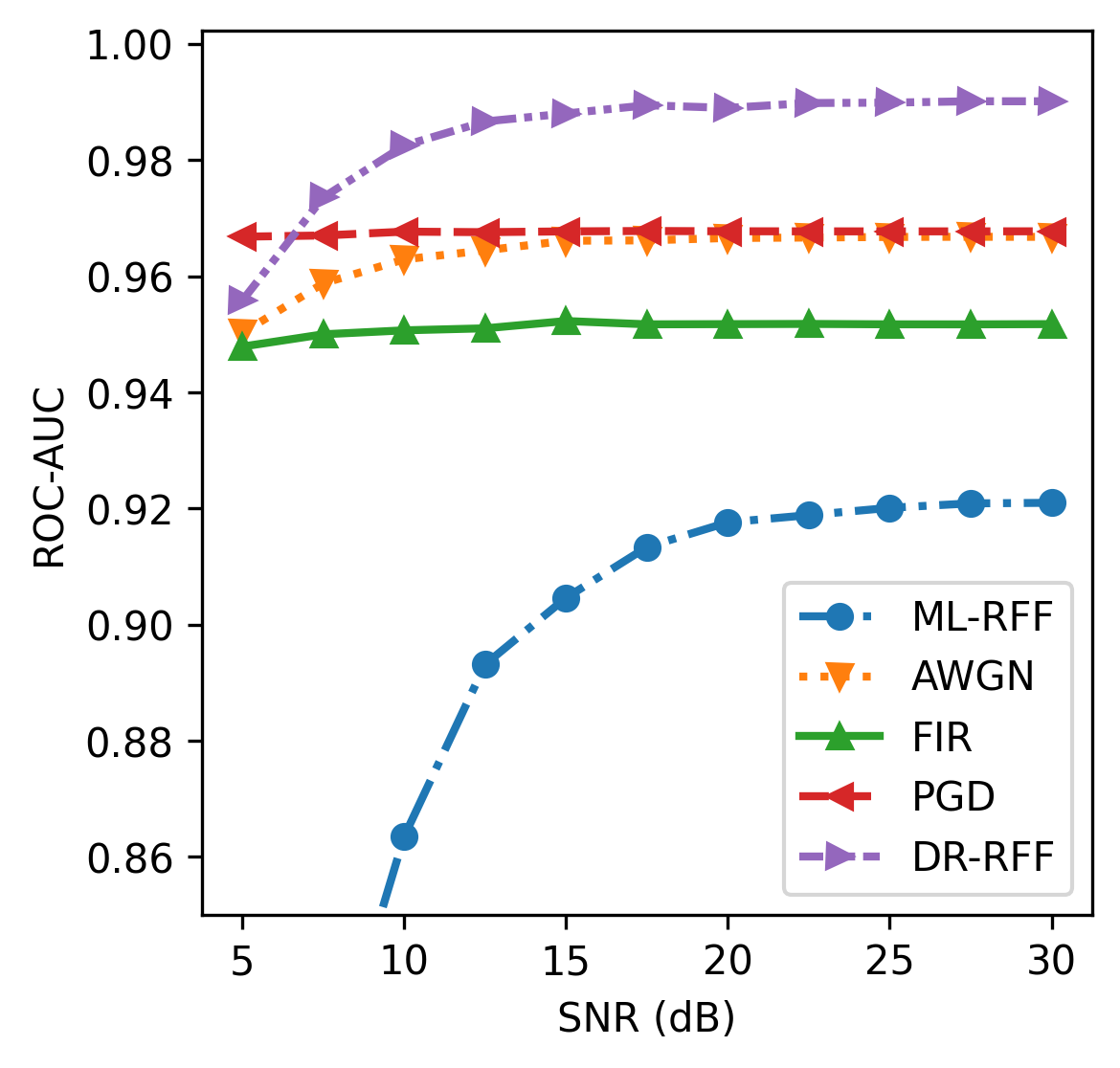}
        \end{minipage}
    }\\
	\caption{\bflag{SNR-AUC curves of different methods under test sets M1-3, which are collected from unknown devices at unknown positions. SNRs of original signals are around 30 dB, and Gaussian noise is added to the signals at steps of 2.5 dB from 5 dB to 27.5~dB. }}
	\label{fig:exp1}
	\vspace{-0.5cm}
\end{figure*}

\subsection{Performance versus SNR}

\bflag{Next, we investigate the robustness of the methods to Gaussian noise. We consider SNRs from 5 to 30 dB by adding Gaussian noises to the unknown channel test sets, M1-M3, and the results are presented in Fig.~\ref{fig:exp1}. }

The results show that \textbf{ML-RFF} has the worst robustness, with performance that degrades  dramatically for test sets when SNR $< 15$ dB. This indicates that the features used in \textbf{ML-RFF} are more sensitive to Gaussian noise than the others. Despite the fact that conventional DA significantly improves the robustness of \textbf{ML-RFF}, there still exists a large gap between these methods and the proposed \textbf{DR-RFF}, especially for the most challenging test set M3, as shown in Fig. \ref{fig:exp1}(c). 
\bflag{Among the conventional DA methods, the discrimination of RFFs from \textbf{AWGN} is better than those from \textbf{FIR} for M2-M3. This is because the assumed channel model in \textbf{AWGN} matches the model of the noise that was added to the data, and therefore is less destructive to its discriminative features than \textbf{FIR}. On the other hand, \textbf{PGA} is less sensitive to both Gaussian noise and real-world channels than handcrafted DA methods, which verify that adversarial training extracts only the most robust features from the training data~\cite{tsipras2018robustness}.}

\bflag{In contrast to these baseline methods, the proposed DR-RFF approach achieves the most discriminative RFF for high SNR situations. It deteriorates as SNR decreases and converges to the same level discrimination as AWGN when SNR $< 7.5$ dB. This reveals that the proposed \textbf{DR-RFF} approach can exploit richer features than conventional DA methods in the training set to improve its discriminability. These additional features are robust to real-world channels but sensitive to Gaussian noise. Even for the low-SNR scenario of the test set M3, the proposed DR-RFF still has a performance edge over the handcrafted DA methods until these additional features are distorted.}
\bflag{These results demonstrate that the proposed DR learning framework can mitigate the mismatch between prior knowledge and real-world situations in the training set.}
\begin{figure*}[t]
	\centering
	\subfigure[Fixed $\alpha=10$ and $\beta=10$.]{
        \centering
        \begin{minipage}[t]{0.308\linewidth}
        \centering        \includegraphics[width=\linewidth]{\rootpath/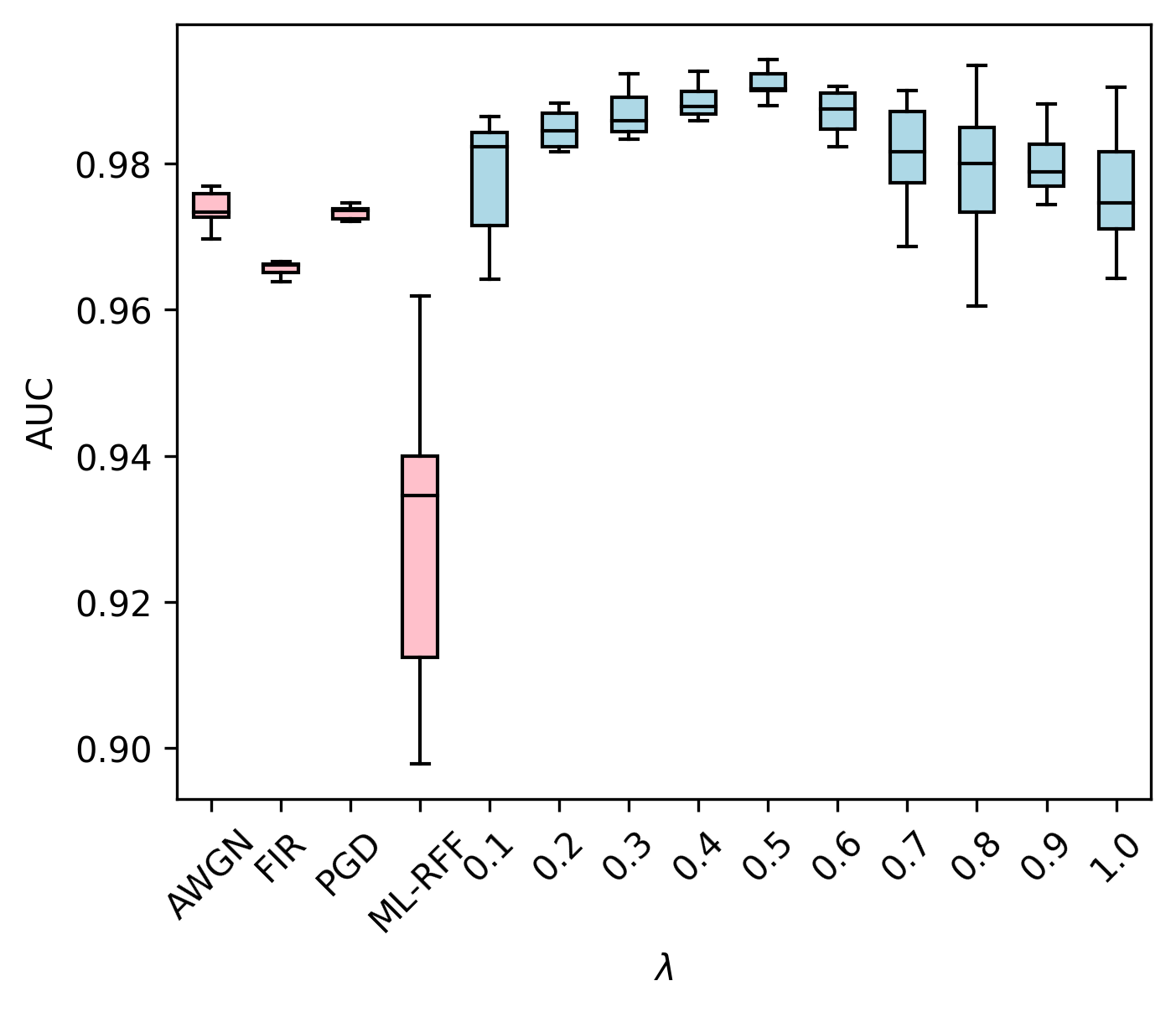}
        \end{minipage}
    }%
	\subfigure[Fixed $\beta=10$ and $\lambda=0.5$.]{
        \centering
        \begin{minipage}[t]{0.32\linewidth}
        \centering        \includegraphics[width=\linewidth]{\rootpath/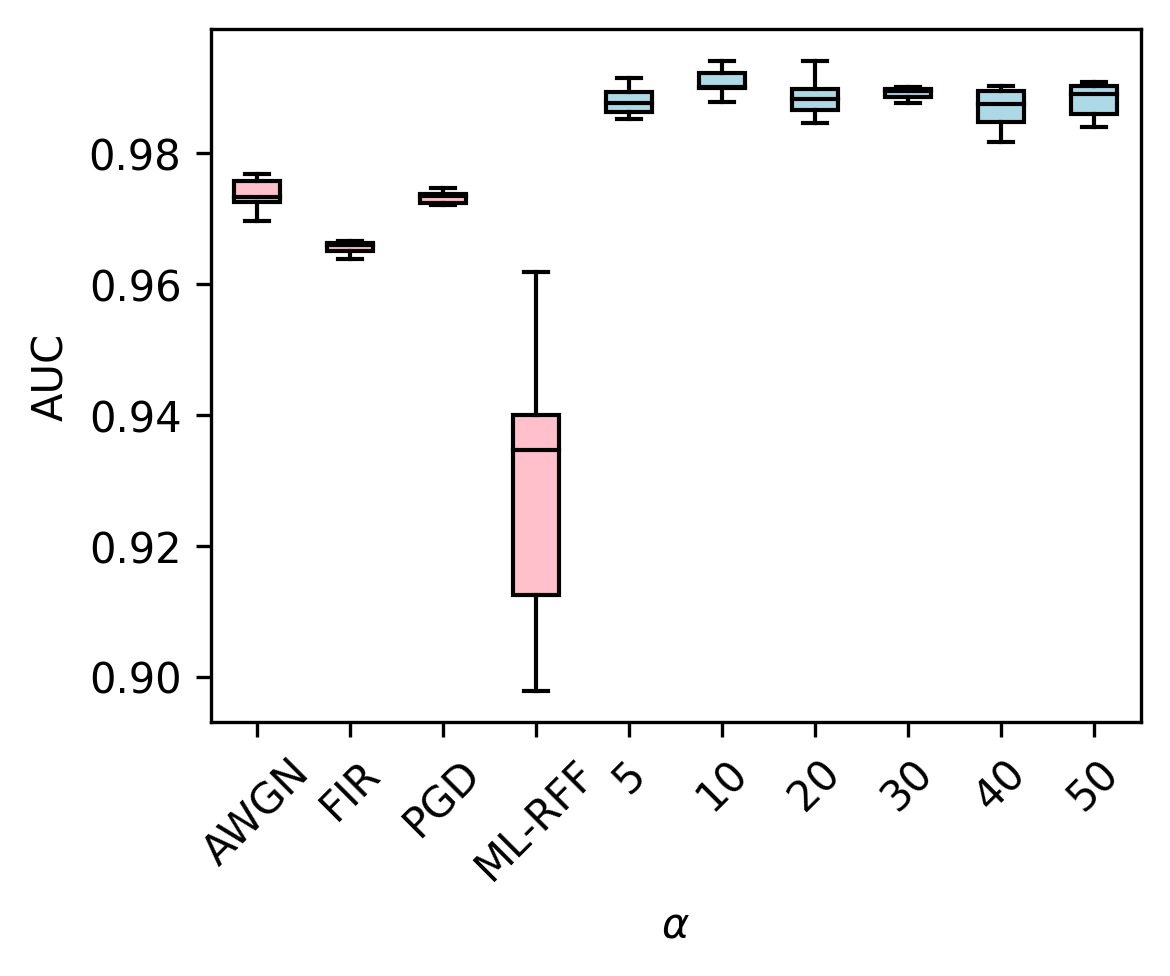}
        \end{minipage}
    }%
	\subfigure[Fixed $\alpha=10$ and $\lambda=0.5$.]{
        \centering
        \begin{minipage}[t]{0.32\linewidth}
        \centering
        \includegraphics[width=\linewidth]{\rootpath/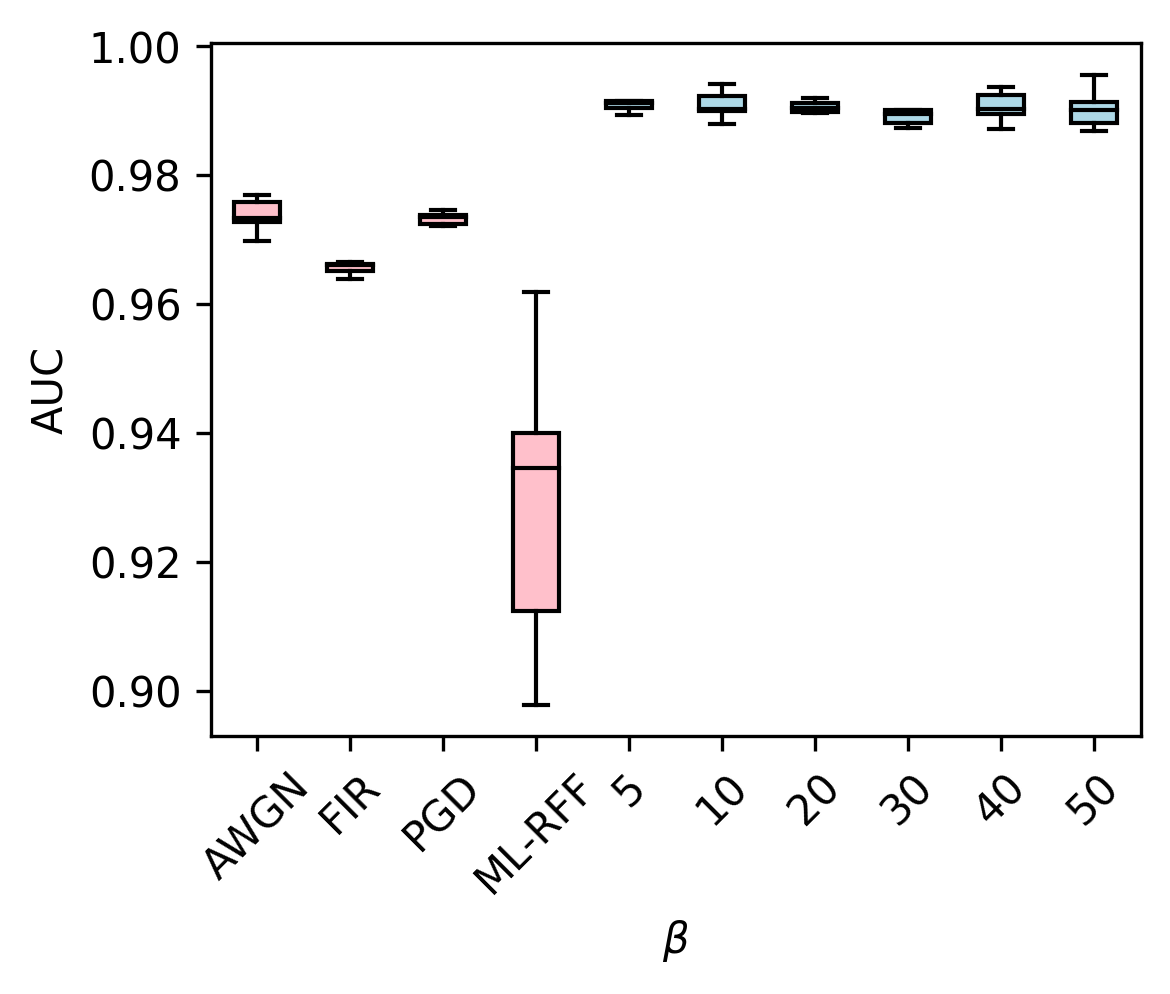}
        \end{minipage}
    }%
	\caption{ \bflag{AUCs on test set M3, respectively achieve by (a) DR-RFFs with different $\lambda$; (b) DR-RFFs with different $\alpha$; and (c) DR-RFFs with different $\beta$. Note that ML-RFF is equivalent to DR-RFF when $\lambda$ is $0$.}}
	\label{fig:params}
	\vspace{-0.5cm}
\end{figure*}

\subsection{Hyper-parameters Tuning}
\bflag{In order to show how $\lambda$, $\alpha$, and $\beta$ affect the performance of the proposed DR learning framework, we include experiments examining the tuning of the hyper-parameters.}
\bflag{We trained each parameter configuration ten times with the other baseline methods and used a box to show the minimum, the maximum, the sample median, and the first and third quartiles of the AUCs corresponding to each configuration of $\lambda$, $\alpha$, and $\beta$. }

\bflag{Fig.~\ref{fig:params}(a) shows that the proposed DR learning framework significantly improves the RFF discrimination for $\lambda \le 0.5$. As the augmented signals gradually dominate the learning effect, i.e., as $\lambda$ grows larger than 0.5, the proposed DR learning framework becomes unstable due to excessive deviations from the raw training dataset. Fig.~\ref{fig:params}(a) also shows that setting hyper-parameter $\alpha \in [0.3, 0.6]$ is a reasonable choice and again leads to the proposed approach significantly outperforming the other baseline methods.} 
\bflag{The box plots in Fig.~\ref{fig:params}(b) and (c) show that when varying $\alpha$ and $\beta$ from 5 to 50, the AUCs of the proposed methods remain nearly unchanged. This implies that the performance of the proposed framework is relatively insensitive to the choice of $\alpha$ and $\beta$ in this range.}


\begin{figure*}[t]
	\centering
	\subfigure[Under known LoS channels~(the validation set)]{
        \centering
        \begin{minipage}[t]{0.45\linewidth}
        \centering
        \includegraphics[width=\linewidth]{\rootpath/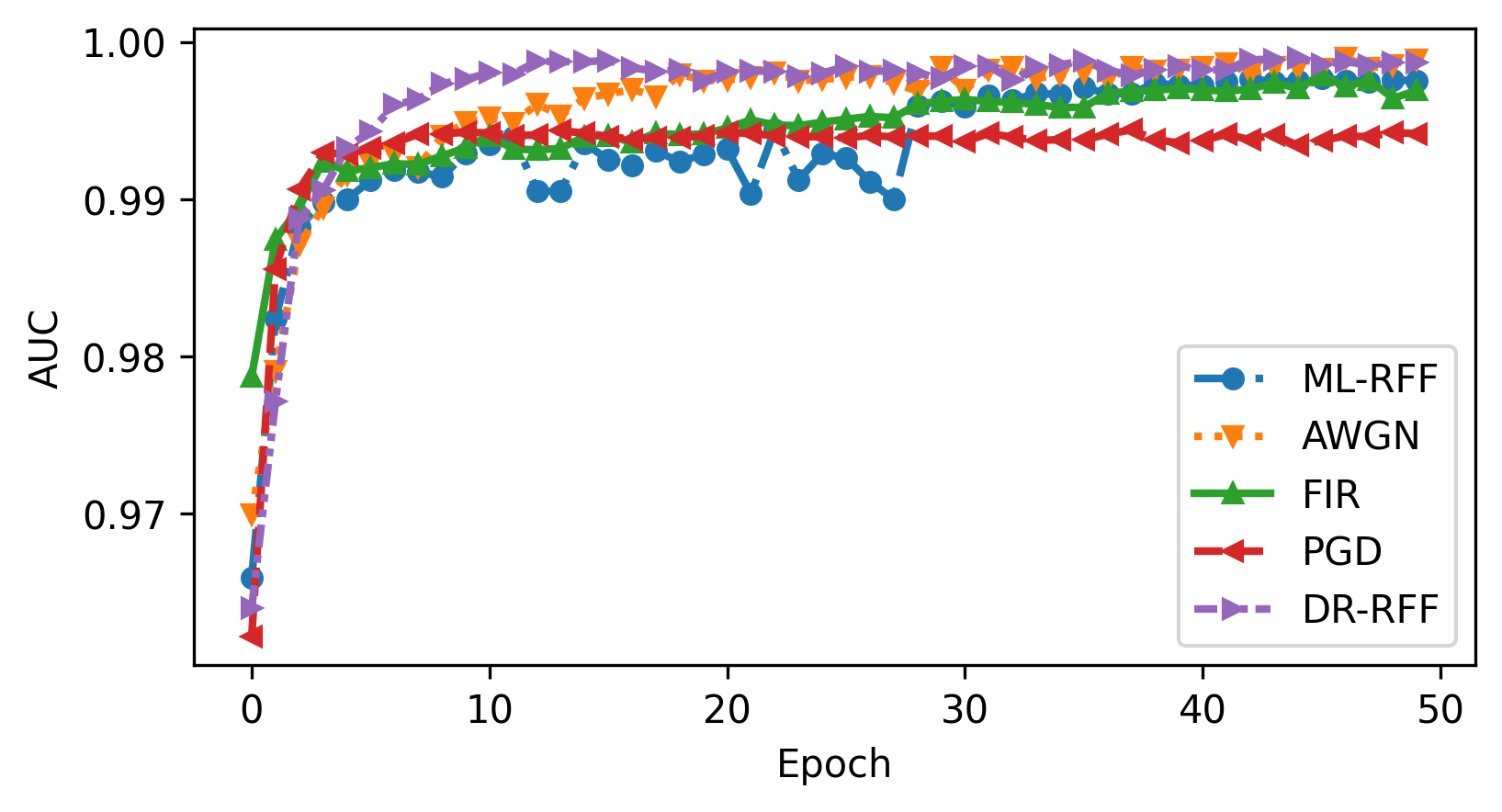}
        \end{minipage}
    }%
    \subfigure[Under unknown multi-path channels~(M3)]{
        \centering
        \begin{minipage}[t]{0.45\linewidth}
        \centering
        \includegraphics[width=\linewidth]{\rootpath/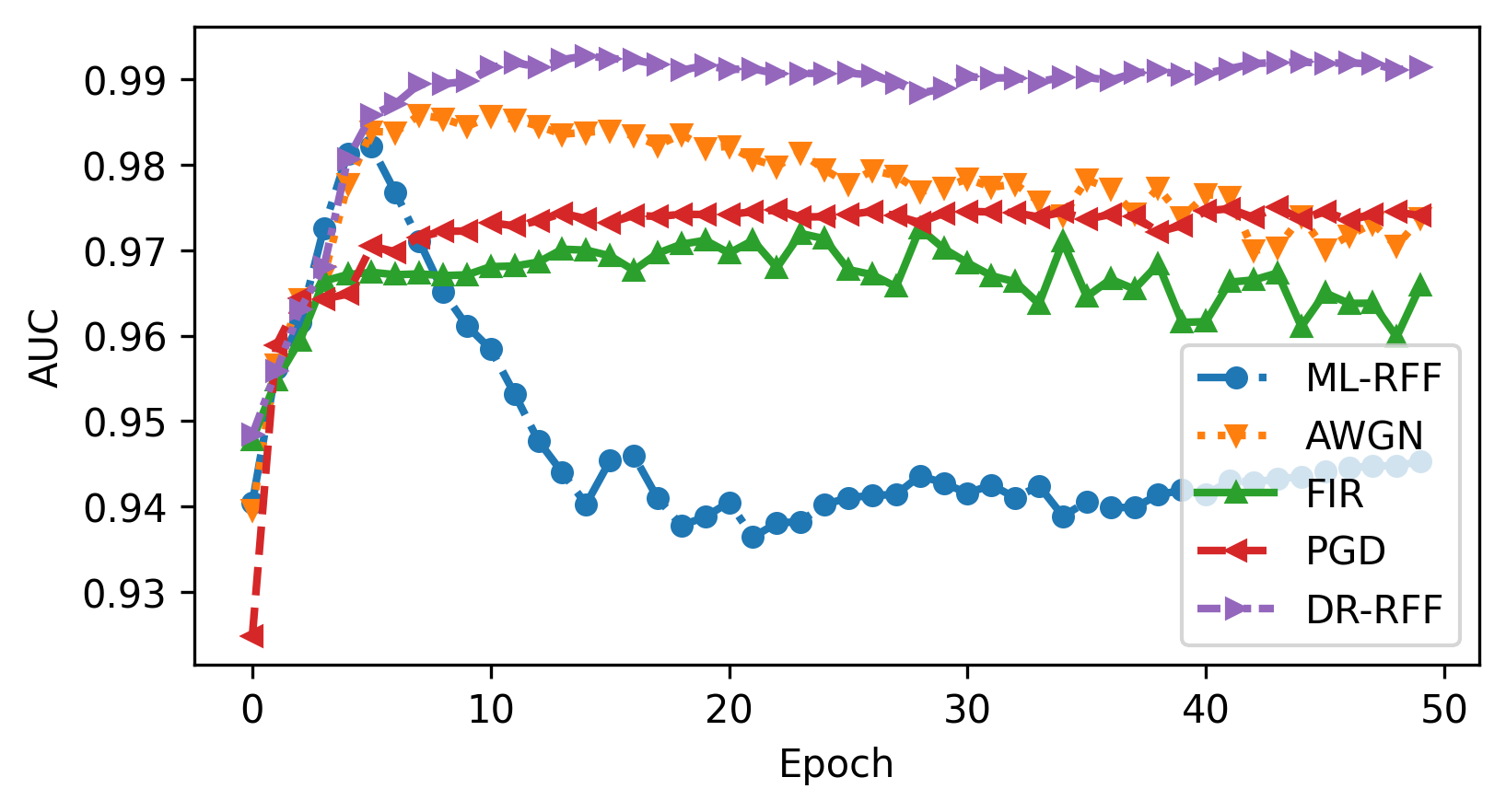}
        \end{minipage}
    }\\
	\caption{\bflag{Average leaning curves of different methods under the validation and open test sets~(SNR $\approx$ 30 dB).}  }
	\label{fig:learning}
	\vspace{-0.5cm}
\end{figure*}

\subsection{Comparison of Learning Curves}
Finally, we compare the proposed \textbf{DR-RFF} with the conventional methods from the perspective of the learning process. We record the performance of the baseline approaches and the proposed \textbf{DR-RFF} in each training epoch and then plot the results in Fig.~\ref{fig:learning}, which shows the average AUC of each method as a function of the training epoch. 
Fig.~\ref{fig:learning}(a) shows the learning curves for the validation set. Since the validation set shares the same data distribution with the training set, the performance achieved here indicates the progress in learning the training set. Fig.~\ref{fig:learning}(b) shows the learning curves for the test set with three types of unknown channels, which reveals potential overfitting to the channel statistics embedded in the training set. 

The learning curves in Fig.~\ref{fig:learning}(a) show that all methods \bflag{except for \textbf{PGD}} can provide a good fit to the training set and near-perfect classification performance under the validation set. However, for the test set with unknown channel statistics in Fig.~\ref{fig:learning}(b), the overfitting of the conventional methods occurs in the early stages of training, e.g., the eighth epoch for \textbf{ML-RFF}. Although handcrafted DA methods can to some extent alleviate the overfitting phenomenon, the degree of overfitting increases as the learning process continues. 

By contrast, overfitting in \textbf{DR-RFF} is suppressed by the proposed DR learning framework. 
\bflag{The proposed framework generates augmented signals based on the current state of the RFF extractor and impose a strong and targeted regularization on the training. In this sense, the proposed framework is a form of adversarial training like \textbf{PGD}, but without impairment of the discrimination. Even towards the end of the training, DR-RFF can adapt to unknown channel statistics and performs well.}
{These results again confirm that applying the proposed DR learning framework can effectively avoid overfitting some channel statistics embedded in the training set.}



\section{Conclusions}
In this paper, we proposed a novel DR learning framework for improving the robustness and generalizability of DL-RFF to unknown channel statistics. DL-RFFs trained using MLE tend to overfit the non-representative channel statistics in the training set and thus lose their generalizability to unknown channels.
To address this problem, we proposed a novel framework that factors the signal into two disjoint parts: a device-relevant representation~(i.e., the RFF) and a device-irrelevant representation~(i.e., the signal background), and can generate signals based on this decomposition. 
Even when all signals in the training set are collected in a simple propagation environment, distinctions in their signal background can still exist. With the help of the proposed framework, we shuffle the signal backgrounds in the training set and mimic transmissions from different types of environments without collecting additional data. In this way, the RFF extractor trained with the proposed framework is encouraged to extract the channel-invariant features as the RFFs.
Our experimental results showed that the proposed framework significantly improved the discriminability of RFFs under unknown multipath fading channels.

\appendices

\section{Detailed Structure of $Q(\cdot, \vecn)$ and $G(\cdot, \cdot)$}
In this section, we introduce the detailed structure of the background signal extractor $Q(\cdot, \vecn)$ and the signal generator $G(\cdot, \cdot)$. Since $Q(\cdot, \vecn)$ and  $G(\cdot, \cdot)$ in this work are both based on U-net~\cite{ronneberger2015u}, we first introduce the basic modules of U-net \cite{ronneberger2015u} and then elaborate on the design of $Q(\cdot, \vecn)$ and $G(\cdot, \cdot)$.
\paragraph{The Basic Modules of U-net} As shown in Table \ref{tb:unet}, U-net consists of three basic modules: \emph{DoubleConv}, \emph{DownConv}, \emph{UpConv}, and \emph{Catenate}. They are described as follows.
\begin{itemize}
	\item \emph{DoubleConv}, contains two convolutional layers with $C_\text{out}$ kernel size of $3\times 3$, 1 padding, 1 stride~(denoted by Conv2D($C_\text{out}$, $3\times 3$, 1, 1)) and the BN+ReLU activation. It takes image $\mathbf{I} \in \mathbb{R}^{C_\text{in}\times W \times H}$ as input, and outputs an image with the same weight and height, i.e., $\mathbf{I} \in \mathbb{R}^{C_\text{out}\times W \times H}$, where $C_\text{in}$, $H$, and $W$ represent the number of channels, the weight, and the height of the image, respectively.
	\item \emph{DownConv}. Down-sampling module, contains one max pooling layer and one DoubleConv module, takes image $\mathbf{I} \in \mathbb{R}^{C_\text{in}\times W \times H}$ as input, and outputs an image with half the weight and height of the input, i.e., $\mathbf{I} \in \mathbb{R}^{C_\text{out}\times \frac{W}{2} \times \frac{H}{2}}$.
	\item \emph{UpConv}. Up-sampling module, contains one up-sampling layer and two convolutional layers, takes image $\mathbf{I} \in \mathbb{R}^{C_\text{in}\times W \times H}$ as input, and outputs an image with twice the weight and height of the input, i.e., $\mathbf{I} \in \mathbb{R}^{C_\text{out}\times \frac{W}{2} \times \frac{H}{2}}$.
	\item \emph{Catenate}. This module merges two images along the channel dimension of the images.
\end{itemize}

\paragraph{Structure of $Q(\cdot, \vecn)$ and $G(\cdot, \cdot)$} As presented in Table~\ref{tb:QG}, both $Q(\cdot, \vecn)$ and $G(\cdot, \cdot)$ contain down-sampling and up-sampling phases. The down-sampling phase consists of one DoubleConv module and four down-sampling modules. The up-sampling phase contains four up-sampling modules, one convolutional layer, and four layers for catenating the outputs from the up-sampling and the down-sampling phases. These catenated layers provide shortcuts that enable the input images to skip the NN processing at different levels of abstraction, thus leading to the property of image domain preservation. The only difference between $Q(\cdot, \vecn)$ and $G(\cdot, \cdot)$ are the layers in the middle. As shown in Table~\ref{tb:QG}, $Q(\cdot, \vecn)$ adds random noise to the latent image, i.e., $\mathbf{I}_5$, while $G(\cdot, \cdot)$ adds the RFF $\vecz$ to $\mathbf{I}_5$.
\ifthenelse{\boolean{isdouble}}{
\begin{table}[!t]  
\caption{The Basic Modules of U-net}
\centering
\resizebox{\linewidth}{!}{
\begin{tabular}{llll}  

\toprule
\multicolumn{2}{l}{{\bf DoubleConv}($C_\text{out}$)}\\
&\multicolumn{3}{l}{{\bf HyperParams}: The number of output channels $C_\text{out}$}\\
\midrule
&\multicolumn{3}{l}{{\bf Input}: Image $\mathbf{I} \in \mathbb{R}^{C_\text{in}\times W \times H}$} \\ 
&\multicolumn{2}{l}{\bf Layers} & {\bf Activation}\\ 
&\multicolumn{2}{l}{1. \quad Conv2D($C_\text{out}$, $3\times3$, 1, 1)} & BN + $\text{ReLU}$\\
&\multicolumn{2}{l}{2. \quad Conv2D($C_\text{out}$, $3\times3$, 1, 1)} & BN + $\text{ReLU}$\\
&\multicolumn{3}{l}{{\bf Output}: Image $\mathbf{I} \in \mathbb{R}^{C_\text{out}\times W \times H}$} \\ 
\midrule 
\midrule
\multicolumn{2}{l}{{\bf DownConv}($C_\text{out}$)}\\
&\multicolumn{3}{l}{{\bf HyperParams}: The number of output channels $C_\text{out}$}\\
\midrule
&\multicolumn{3}{l}{{\bf Input}: Image $\mathbf{I} \in \mathbb{R}^{C_\text{in}\times W \times H}$} \\ 
&\multicolumn{2}{l}{\bf Layers} & \\ 
&\multicolumn{2}{l}{1. \quad MaxPool2d(2)}\\
&\multicolumn{2}{l}{2. \quad DoubleConv($C_\text{out}$)} \\
&\multicolumn{3}{l}{{\bf Output}: Image $\mathbf{I} \in \mathbb{R}^{C_\text{out}\times \frac{W}{2} \times \frac{H}{2}}$} \\ 
\midrule 
\midrule
\multicolumn{2}{l}{{\bf UpConv}($C_\text{out}$)}\\
&\multicolumn{3}{l}{{\bf HyperParams}: The number of output channels $C_\text{out}$}\\
\midrule
&\multicolumn{3}{l}{{\bf Input}: Image $\mathbf{I} \in \mathbb{R}^{C_\text{in}\times W \times H}$} \\ 
&\multicolumn{2}{l}{\bf Layers} & {\bf Activation}\\ 
&\multicolumn{2}{l}{1. \quad Upsample(2)}\\
&\multicolumn{2}{l}{2. \quad Conv2D($C_\text{in}/2$, $3\times3$, 1, 1)} & BN + $\text{ReLU}$\\
&\multicolumn{2}{l}{3. \quad Conv2D($C_\text{out}$, $3\times3$, 1, 1)} & BN + $\text{ReLU}$\\
&\multicolumn{3}{l}{{\bf Output}: Image $\mathbf{I} \in \mathbb{R}^{C_\text{out}\times 2W \times 2H}$} \\ 
\midrule
\midrule
\multicolumn{2}{l}{\bf Catenate}\\
&\multicolumn{3}{l}{Catenating two images along the dimension of channels.} \\ 
\midrule
&\multicolumn{3}{l}{{\bf Input}: Images $\mathbf{I}_1 \in \mathbb{R}^{C_1\times W \times H}$, and $\mathbf{I}_2 \in \mathbb{R}^{C_2\times W \times H}$} \\ 
&\multicolumn{3}{l}{{\bf Output}: Image $\mathbf{I} \in \mathbb{R}^{(C_1 + C_2)\times W \times H}$}\\ 
\bottomrule  
\end{tabular}
}
\label{tb:unet}
\end{table}
\begin{table}[!t]  

\caption{The Structure of $Q(\cdot, \vecn)$ or $G(\cdot, \cdot)$}
\centering
\resizebox{\linewidth}{!}{
\begin{tabular}{llll}  

\toprule
\multicolumn{4}{l}{{\bf Input}: Signal $\vecx \in \mathbb{C}^{M}$ $\rightarrow$ Image $\mathbf{I} \in \mathbb{R}^{2\times \frac{M}{S} \times S}$} \\  

\midrule  
{\bf Layers} &{\bf Inputs $\rightarrow$} & {\bf Modules $\longrightarrow$} & {\bf Outputs}\\
\midrule  
\multicolumn{4}{l}{{\# Down-sampling phase:}} \\  
1 & $\mathbf{I}$    & DoubleConv(64) &  $\mathbf{I}_1 $\\
2 & $\mathbf{I}_1$  & DownConv(128)   &  $\mathbf{I}_2  $\\ 
3 & $\mathbf{I}_2$  & DownConv(256)  &  $\mathbf{I}_3 $\\ 
4 &  $\mathbf{I}_3$  & DownConv(512)  &  $\mathbf{I}_4 $\\ 
5 & $\mathbf{I}_4$  & DownConv(512)  &  $\mathbf{I}_5 $\\ 
\midrule 
\multicolumn{4}{l}{{\# {\bf For $Q(\cdot, \vecn)$}: Adding randomness with $\vecn \sim \mathcal{N}(\mathbf{0}, \mathbf{I})$}}\\
6 & $\mathbf{I}_5$ and $\vecn$ & {$\mathbf{I}_5^* = \mathbf{I}_5 + \vecn$} & $\mathbf{I}_5^*$\\
\midrule 
\multicolumn{4}{l}{{\# {\bf For $G(\cdot, \cdot)$}: Adding RFF}}\\
6-1 & $\vecz$  & FC(the shape of $\mathbf{I}_5$)  &  $\mathbf{I}_\vecz $\\ 
6-2 & $\mathbf{I}_5$ and $\mathbf{I}_\vecz$ & {$\mathbf{I}_5^* = \mathbf{I}_5 + \mathbf{I}_\vecz$} & $\mathbf{I}_5^*$\\
\midrule 
\multicolumn{4}{l}{{\# Up-sampling phase:}} \\ 
7 & $\mathbf{I}_5^*$ & UpConv(256)  &  $\mathbf{I}_4^\prime $\\
8 & $\mathbf{I}_4$ and $\mathbf{I}_4^\prime$ & Catenate     &  $\mathbf{I}_4^*$\\
9 & $\mathbf{I}_4^*$ & UpConv(128)  &  $\mathbf{I}_3^\prime$\\ 
10 & $\mathbf{I}_3$ and $\mathbf{I}_3^\prime$ & Catenate    &  $\mathbf{I}_3^*$\\
11 & $\mathbf{I}_3^*$ & UpConv(64)   &  $\mathbf{I}_2^\prime$\\ 
12 & $\mathbf{I}_2$ and $\mathbf{I}_2^\prime$ & Catenate    &  $\mathbf{I}_2^*$\\
13 & $\mathbf{I}_2^*$ & UpConv(64)  & $\mathbf{I}_1^\prime$\\  
14 & $\mathbf{I}_1$ and $\mathbf{I}_1^\prime$ & Catenate    &  $\mathbf{I}_1^*$\\
15 & $\mathbf{I}_1^*$ &  Conv2D(2, $1\times1$, 1, 1) &$\mathbf{I}_\text{out} $\\ 
%
\midrule
\multicolumn{4}{l}{{\bf Output}: Image $\mathbf{I}_\text{out} \in \mathbb{R}^{2\times \frac{M}{S} \times S}$ $\rightarrow$ Signal $\vecx_\text{out} \in \mathbb{C}^{M}$ }\\
\bottomrule  

\end{tabular}
}
\label{tb:QG}
\end{table}
}{
\begin{table}[!t]  
\begin{minipage}[t]{0.49\textwidth}
\caption{The Basic Modules of U-net}
\centering
\resizebox{\linewidth}{!}{
\begin{tabular}{llll}  

\toprule
\multicolumn{2}{l}{{\bf DoubleConv}($C_\text{out}$)}\\
&\multicolumn{3}{l}{{\bf HyperParams}: The number of output channels $C_\text{out}$}\\
\midrule
&\multicolumn{3}{l}{{\bf Input}: Image $\mathbf{I} \in \mathbb{R}^{C_\text{in}\times W \times H}$} \\ 
&\multicolumn{2}{l}{\bf Layers} & {\bf Activation}\\ 
&\multicolumn{2}{l}{1. \quad Conv2D($C_\text{out}$, $3\times3$, 1, 1)} & BN + $\text{ReLU}$\\
&\multicolumn{2}{l}{2. \quad Conv2D($C_\text{out}$, $3\times3$, 1, 1)} & BN + $\text{ReLU}$\\
&\multicolumn{3}{l}{{\bf Output}: Image $\mathbf{I} \in \mathbb{R}^{C_\text{out}\times W \times H}$} \\ 
\midrule 
\midrule
\multicolumn{2}{l}{{\bf DownConv}($C_\text{out}$)}\\
&\multicolumn{3}{l}{{\bf HyperParams}: The number of output channels $C_\text{out}$}\\
\midrule
&\multicolumn{3}{l}{{\bf Input}: Image $\mathbf{I} \in \mathbb{R}^{C_\text{in}\times W \times H}$} \\ 
&\multicolumn{2}{l}{\bf Layers} & \\ 
&\multicolumn{2}{l}{1. \quad MaxPool2d(2)}\\
&\multicolumn{2}{l}{2. \quad DoubleConv($C_\text{out}$)} \\
&\multicolumn{3}{l}{{\bf Output}: Image $\mathbf{I} \in \mathbb{R}^{C_\text{out}\times \frac{W}{2} \times \frac{H}{2}}$} \\ 
\midrule 
\midrule
\multicolumn{2}{l}{{\bf UpConv}($C_\text{out}$)}\\
&\multicolumn{3}{l}{{\bf HyperParams}: The number of output channels $C_\text{out}$}\\
\midrule
&\multicolumn{3}{l}{{\bf Input}: Image $\mathbf{I} \in \mathbb{R}^{C_\text{in}\times W \times H}$} \\ 
&\multicolumn{2}{l}{\bf Layers} & {\bf Activation}\\ 
&\multicolumn{2}{l}{1. \quad Upsample(2)}\\
&\multicolumn{2}{l}{2. \quad Conv2D($C_\text{in}/2$, $3\times3$, 1, 1)} & BN + $\text{ReLU}$\\
&\multicolumn{2}{l}{3. \quad Conv2D($C_\text{out}$, $3\times3$, 1, 1)} & BN + $\text{ReLU}$\\
&\multicolumn{3}{l}{{\bf Output}: Image $\mathbf{I} \in \mathbb{R}^{C_\text{out}\times 2W \times 2H}$} \\ 
\midrule
\midrule
\multicolumn{2}{l}{\bf Catenate}\\
&\multicolumn{3}{l}{Catenating two images along the dimension of channels.} \\ 
\midrule
&\multicolumn{3}{l}{{\bf Input}: Images $\mathbf{I}_1 \in \mathbb{R}^{C_1\times W \times H}$, and $\mathbf{I}_2 \in \mathbb{R}^{C_2\times W \times H}$} \\ 
&\multicolumn{3}{l}{{\bf Output}: Image $\mathbf{I} \in \mathbb{R}^{(C_1 + C_2)\times W \times H}$}\\ 
\bottomrule  
\end{tabular}
}
\label{tb:unet}
\end{minipage}
\begin{minipage}[t]{0.49\textwidth}
	\caption{The Structure of $Q(\cdot, \vecn)$ or $G(\cdot, \cdot)$}
\centering
\resizebox{\linewidth}{!}{
\begin{tabular}{llll}  

\toprule
\multicolumn{4}{l}{{\bf Input}: Signal $\vecx \in \mathbb{C}^{M}$ $\rightarrow$ Image $\mathbf{I} \in \mathbb{R}^{2\times \frac{M}{S} \times S}$} \\  

\midrule  
{\bf Layers} &{\bf Inputs $\rightarrow$} & {\bf Modules $\longrightarrow$} & {\bf Outputs}\\
\midrule  
\multicolumn{4}{l}{{\# Down-sampling phase:}} \\  
1 & $\mathbf{I}$    & DoubleConv(64) &  $\mathbf{I}_1 $\\
2 & $\mathbf{I}_1$  & DownConv(128)   &  $\mathbf{I}_2  $\\ 
3 & $\mathbf{I}_2$  & DownConv(256)  &  $\mathbf{I}_3 $\\ 
4 &  $\mathbf{I}_3$  & DownConv(512)  &  $\mathbf{I}_4 $\\ 
5 & $\mathbf{I}_4$  & DownConv(512)  &  $\mathbf{I}_5 $\\ 
\midrule 
\multicolumn{4}{l}{{\# {\bf For $Q(\cdot, \vecn)$}: Adding randomness with $\vecn \sim \mathcal{N}(\mathbf{0}, \mathbf{I})$}}\\
6 & $\mathbf{I}_5$ and $\vecn$ & {$\mathbf{I}_5^* = \mathbf{I}_5 + \vecn$} & $\mathbf{I}_5^*$\\
\midrule 
\multicolumn{4}{l}{{\# {\bf For $G(\cdot, \cdot)$}: Adding RFF}}\\
6-1 & $\vecz$  & FC(the shape of $\mathbf{I}_5$)  &  $\mathbf{I}_\vecz $\\ 
6-2 & $\mathbf{I}_5$ and $\mathbf{I}_\vecz$ & {$\mathbf{I}_5^* = \mathbf{I}_5 + \mathbf{I}_\vecz$} & $\mathbf{I}_5^*$\\
\midrule 
\multicolumn{4}{l}{{\# Up-sampling phase:}} \\ 
7 & $\mathbf{I}_5^*$ & UpConv(256)  &  $\mathbf{I}_4^\prime $\\
8 & $\mathbf{I}_4$ and $\mathbf{I}_4^\prime$ & Catenate     &  $\mathbf{I}_4^*$\\
9 & $\mathbf{I}_4^*$ & UpConv(128)  &  $\mathbf{I}_3^\prime$\\ 
10 & $\mathbf{I}_3$ and $\mathbf{I}_3^\prime$ & Catenate    &  $\mathbf{I}_3^*$\\
11 & $\mathbf{I}_3^*$ & UpConv(64)   &  $\mathbf{I}_2^\prime$\\ 
12 & $\mathbf{I}_2$ and $\mathbf{I}_2^\prime$ & Catenate    &  $\mathbf{I}_2^*$\\
13 & $\mathbf{I}_2^*$ & UpConv(64)  & $\mathbf{I}_1^\prime$\\  
14 & $\mathbf{I}_1$ and $\mathbf{I}_1^\prime$ & Catenate    &  $\mathbf{I}_1^*$\\
15 & $\mathbf{I}_1^*$ &  Conv2D(2, $1\times1$, 1, 1) &$\mathbf{I}_\text{out} $\\ 
%
\midrule
\multicolumn{4}{l}{{\bf Output}: Image $\mathbf{I}_\text{out} \in \mathbb{R}^{2\times \frac{M}{S} \times S}$ $\rightarrow$ Signal $\vecx_\text{out} \in \mathbb{C}^{M}$ }\\
\bottomrule  

\end{tabular}
}
\label{tb:QG}
\end{minipage}
\vspace{-0.5cm}
\end{table}
}
\ifCLASSOPTIONcaptionsoff
  \newpage
\fi





%




\end{document}